\pdfoutput=1
\documentclass[prb,twocolumn,showpacs,preprintnumbers,superscriptaddress,amsmath,amssymb]{revtex4-2}
\usepackage{graphicx}
\usepackage{float}
\usepackage{dcolumn}
\usepackage{color}
\usepackage[dvipsnames]{xcolor} 
\usepackage{latexsym,bm}
\usepackage[normalem]{ulem}
\usepackage{multirow}
\usepackage{appendix}
\usepackage{amsmath}
\usepackage{amsfonts}
\usepackage{booktabs}
\usepackage{subcaption}
\usepackage{tabularx}

\usepackage{threeparttable} 

\newcommand{\MC}[1]{\textcolor{black}{{#1}}}
\newcommand{\SL}[1]{\textcolor{black}{{#1}}}

\newcommand{\tx}[1]{\text{#1}}

\begin{document}

\hyphenpenalty=5000
\tolerance=1000

\title{Machine learning based nonlocal kinetic energy density functional for simple metals and alloys}

\author{Liang Sun}
\affiliation{HEDPS, CAPT, School of Physics and College of Engineering, Peking University, Beijing 100871, P. R. China}
\author{Mohan Chen}
\email{mohanchen@pku.edu.cn}
\affiliation{HEDPS, CAPT, School of Physics and College of Engineering, Peking University, Beijing 100871, P. R. China}
\affiliation{AI for Science Institute, Beijing 100080, P. R. China}
\date{\today}

\begin{abstract}
Developing an accurate kinetic energy density functional (KEDF) remains a major hurdle in orbital-free density functional theory.
We propose a machine learning based physical-constrained nonlocal (MPN) KEDF \MC{and implement it with the usage of the bulk-derived local pseudopotentials and plane wave basis sets in the ABACUS package. The MPN KEDF is designed to satisfy} three exact physical constraints: the scaling law of electron kinetic energy, the free electron gas limit, and the non-negativity of Pauli energy density.
The MPN KEDF is systematically tested for simple metals, including Li, Mg, Al, and 59 alloys.
We conclude that incorporating nonlocal information for designing new KEDFs and obeying exact physical constraints are essential to improve the accuracy, transferability, and stability of ML-based KEDF.
These results shed new light on the construction of ML-based functionals.
\end{abstract}
\maketitle

\section{Introduction}
Kohn-Sham density functional theory (KSDFT) is a widely-used {\it ab initio} method in materials science.~\cite{64PR-Hohenberg, 65PR-Kohn} However, its computational complexity of $O(N^3)$, where $N$ is the number of atoms, poses significant challenges for large systems.
Alternatively, orbital-free density functional theory (OFDFT)~\cite{02Carter,18JMR-Witt} calculates the non-interacting electron kinetic energy $T_s$ directly from the charge density instead of relying on the one-electron Kohn-Sham orbitals.
As a result, OFDFT achieves a more affordable computational complexity of typically $O(N\ln{N})$ or $O(N)$.~\cite{08CPC-Ho-PROFESS, 10CPC-Hung-PROFESS, 15CPC-Chen-PROFESS, 16CPC-Mi-atlas}
%
%
Given that $T_s$ is comparable in magnitude to the total energy,
the accuracy of OFDFT largely depends on the approximated form of the kinetic energy density functional (KEDF). However, developing an accurate KEDF has been a major hurdle in the field of OFDFT for decades years.

Over the past few decades, continuous efforts have been devoted to developing analytical KEDFs.~\cite{12CPC-Karasiev, 18JMR-Witt}
In general, KEDFs can be classified into two categories. 
The first category comprises local and semilocal components in KEDFs, where the kinetic energy density is a function of the charge density, the charge density gradient, the Laplacian of charge density, or even higher-order derivatives of the charge density.~\cite{27-Thomas-local, 27TANL-Fermi-local, 35-vW-semilocal, 18B-Luo-semilocal, 18JPCL-Constantin-semilocal, 20Kang-semilocal}
%
%
%
The second category consists of nonlocal forms of KEDFs, where the kinetic energy density is a functional of charge density, such that the kinetic energy density at each point in real space depends on the nonlocal charge density.~\cite{92B-Wang-nonlocal, 99B-Wang-nonlocal, 10B-Huang-nonlocal, 18JCP-Mi-nonlocal, 21B-Shao-nonlocal}
Typically, semilocal KEDFs are more computationally efficient, while nonlocal ones offer a higher accuracy. 
However, since most of the existing nonlocal KEDFs are constructed based on the Lindhard response function, which is accurate for nearly free electron gas, they are mainly adequate for simple metals.~\cite{92B-Wang-nonlocal, 99B-Wang-nonlocal}
Some KEDFs were proposed to describe semiconductor systems, but they cannot work well for simple metals.~\cite{10B-Huang-nonlocal, 18JCP-Mi-nonlocal, 21B-Shao-nonlocal}
As a result, a KEDF that works for both simple metal and semiconductor systems is still lacking, and it is still unclear how to construct it systematically.

In recent years, machine learning (ML) techniques have been involved in the developments of computational physics.~\cite{23Science-Huang-mlfp}
In particular, the remarkable fitting ability of ML models has been demonstrated in various applications, including the fitting of potential energy surfaces in molecular dynamics~\cite{18prl-zhang-dp, 20CPC-Zhang-dpgen}, as well as fitting exchange-correlation functionals~\cite{20JCTC-chen-deepks, 21L-Kasim-mlxc, 21s-kirkpatrick-dm21, 22PRR-nagai-mlxc} and Hamiltonian matrices~\cite{22NCS-Li-deepH} within the framework of density functional theory (DFT).~\cite{64PR-Hohenberg,65PR-Kohn}
Additionally, there have been endeavors to construct ML-based KEDFs within the framework of OFDFT.~\cite{12L-Snyder-mlof, 18TJCP-Seino-mlof, 18TJCP-Hollingsworth-mlof, 20JCTC-Meyer-mlof, 21PRR-Imoto-mlof, 22JCTC-Ryczko-mlof, 23JCTC-Pablo}
For example, Imoto \emph{et al.} implemented a semilocal ML-based KEDF, which takes dimensionless gradient and dimensionless Laplacian of charge density as descriptors and puts the enhancement factor of kinetic energy density as the output of neural network (NN).~\cite{21PRR-Imoto-mlof}
This model exhibits convergence and satisfies the scaling law, but it overlooks nonlocal information crucial for improving the accuracy of KEDFs.
Ryczko \emph{et al.}. implemented a nonlocal ML-based KEDF, utilizing a voxel deep NN, but this model could not achieve convergence in OFDFT computations.~\cite{22JCTC-Ryczko-mlof}
Thus, it is still a formidable task to construct an accurate, transferable, and computationally stable ML-based KEDF.

\begin{figure}[thbp]
	\centering
	\includegraphics[width=\linewidth]{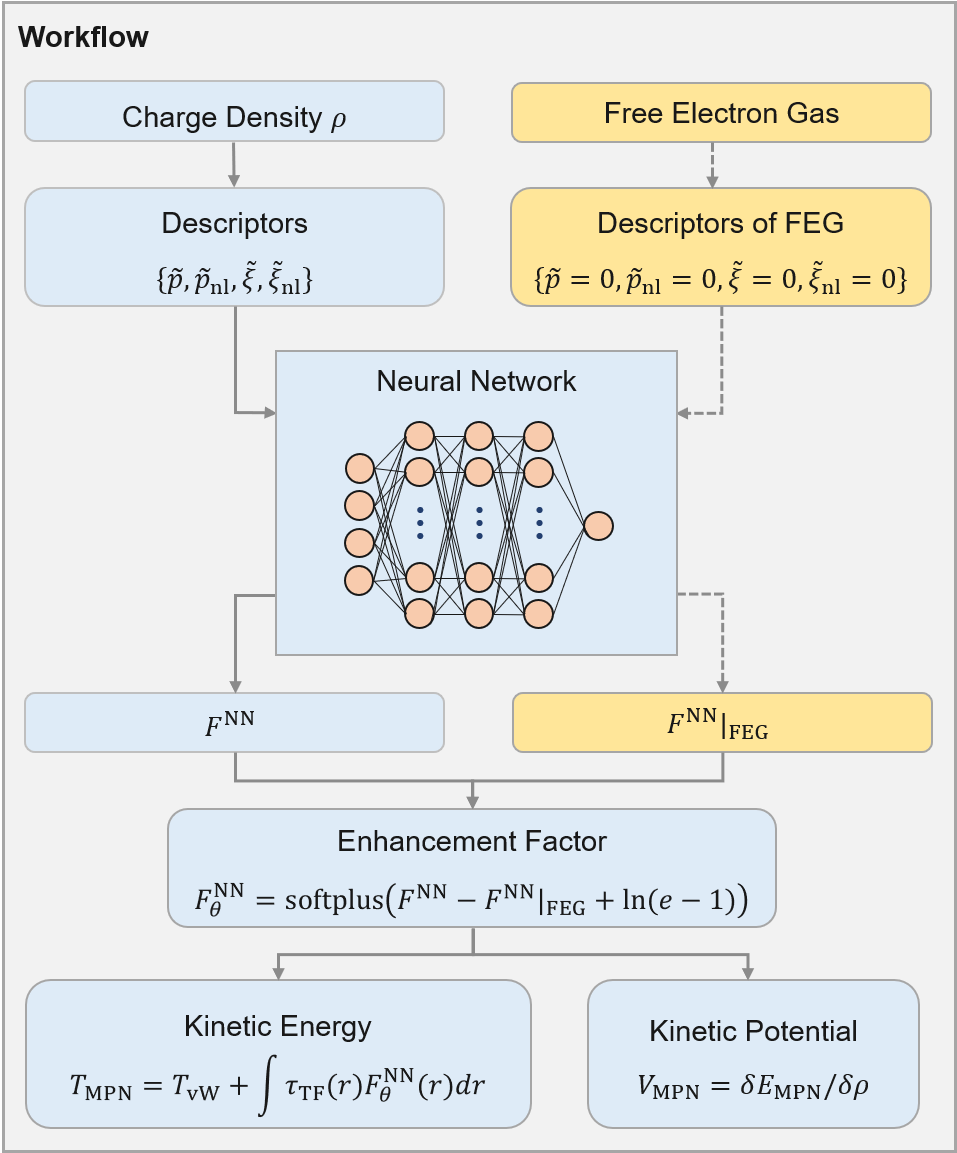}\\
	\caption{
 \MC{
Workflow of the MPN KEDF. 
$F^{\rm{NN}}(r)$ is the enhancement factor obtained by the deep neural network (NN), and $F^{\rm{NN}}|_{\rm{FEG}}$ denotes the enhancement factor under the free electron gas (FEG) limit.
In order to ensure both the FEG limit and the non-negativity of Pauli energy density are satisfied, the enhancement factor of Pauli energy is defined as $F_{\rm{\theta}}^{\rm{NN}} = {\rm{softplus}}\left(F^{\rm{NN}} - F^{\rm{NN}}|_{\rm{FEG}} + \ln{(e-1)}\right)$, where ${\rm{softplus}}(x)=\ln(1+e^x)$ is an activation function commonly used in machine learning with ${\rm{softplus}}(x)|_{x=\ln(e-1)}=1$. 
The defined formulas are used to evaluate the kinetic energy and kinetic potential.}
	}\label{fig:workflow}
\end{figure}

\begin{figure}[htbp]
	\centering
	\includegraphics[width=0.95\linewidth]{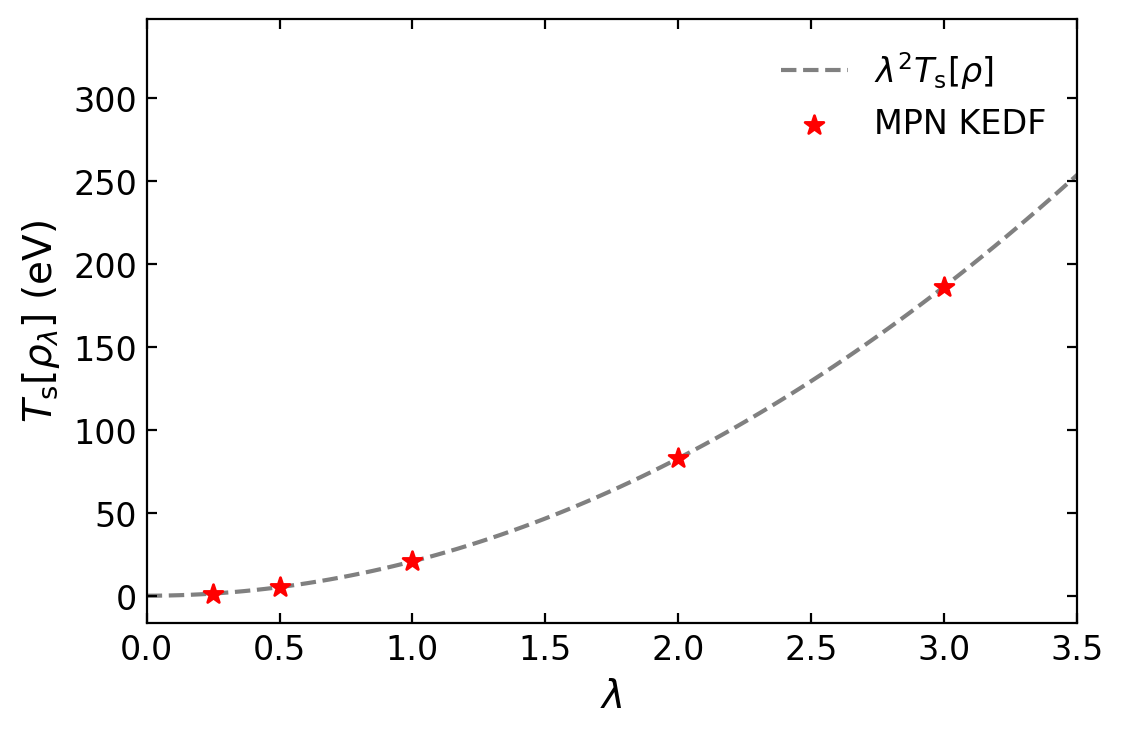}\\
	\caption{
Illustration of the scaling law introduced in Eq.~\ref{eq.scaling}. The gray line represents the function of $\lambda^2 T_{\rm{s}}[\rho]$, where $\rho$ denotes the ground charge density of face-centered cubic (fcc) Al as obtained by the MPN KEDF. 
 The red stars denote the kinetic energies of $\rho_{\lambda}=\lambda^3\rho(\lambda r)$  computed using the MPN KEDF for different values of $\lambda$, namely 0.25, 0.5, 1.0, 2.0, and 3.0.
 All the red stars fall on the gray line, indicating that the scaling law $T_{\rm{s}}[\rho_{\lambda}] = \lambda^2 T_{\rm{s}}[\rho]$ is exactly obeyed by the MPN KEDF.
	}\label{fig:Scaling}
\end{figure}

In this work, as the first step to construct an ML-based KEDF that works for both simple metal and semiconductor systems, we construct an ML-based physical-constrained nonlocal KEDF (MPN KEDF) for simple metals and their alloys, which (a) contains nonlocal information, (b) obeys a series of exact physical constraints, and (c) achieves convergence via careful design of descriptors, NN output, post-processing, and loss function, etc. 
The performance of the MPN KEDF is systematically evaluated by testing on a series of simple metals, including lithium (Li), magnesium (Mg), aluminum (Al), and their alloys.
In particular, incorporating nonlocal information and exact physical constraints is crucial to improving the accuracy, transferability, and stability of ML-based KEDFs.~\cite{18TJCP-Hollingsworth-mlof}

The rest of this paper is organized as follows.
In Section II, we propose an ML-based KEDF that satisfies physical constraints and introduces numerical details of KSDFT and OFDFT calculations.
In Section III, we analyze the performances of the MPN KEDF and discuss the results.
Finally, the conclusions are drawn in Section IV.

\section{Methods}
\subsection{Pauli Energy and Pauli Potential}

In general, the non-interacting kinetic energy $T_s$ can be divided into two parts,~\cite{88PRA-Levy-pauli}
\begin{equation}
    T_{{s}} = T_{\rm{vW}} + T_{\rm{\theta}},
\end{equation}
where 
\begin{equation}
T_{\rm{vW}} = \frac{1}{8} \int {\frac{{\left| {\nabla \rho ({r})} \right|}^2}{\rho \,({r})} \,\tx{d}^3 {r}}
\end{equation}
is the von Weizs$\mathrm{\Ddot{a}}$cker (vW) KEDF,~\cite{35-vW-semilocal} a rigorous lower bound to the $T_s$, with $\rho(r)$ being the charge density. 
The second term $T_{\rm{\theta}}$ represents the Pauli energy, which takes the form of
\begin{equation}
    T_{\rm{\theta}} = \int{\tau_{\rm{TF}} F_{\rm{\theta}} {\rm{d}}^3 {r}},
\end{equation}
where the Thomas-Fermi (TF) kinetic energy density~\cite{27-Thomas-local, 27TANL-Fermi-local} term is
\begin{equation}
\tau_{\rm{TF}} = \frac{3}{10}(3\pi^2)^{2/3} \rho^{5/3}.
\end{equation}
Additionally, $F_{\rm{\theta}}$ denotes the enhancement factor.
The corresponding Pauli potential is given by 
\begin{equation}
V_{\rm{\theta}}(r) = \delta E_{\rm{\theta}}/\delta \rho(r).
\end{equation}
The Pauli energy and Pauli potential satisfy several exact physical constraints. For example, first, the scaling law is
\begin{equation}
T_{\rm{\theta}}[\rho_{\lambda}] = \lambda^2 T_{\rm{\theta}}[\rho], 
\label{eq.scaling}
\end{equation}
where $\rho_{\lambda}=\lambda^3\rho(\lambda r)$ and $\lambda$ is a positive number.~\cite{88PRA-Levy-pauli}

Second, in the free electron gas (FEG) limit, the TF KEDF is exact, and the vW part vanishes so that the enhancement factor in the FEG limit takes the form of
\begin{equation}
F_{\rm{\theta}}(r)|_{\rm{FEG}} = 1.
\label{eq.feg_f}
\end{equation}
In addition, the Pauli potential returns to the potential of TF KEDF $V_{\rm{TF}}(r)$
\begin{equation}
V_{\rm{\theta}}(r)|_{\rm{FEG}} = V_{\rm{TF}}(r) = \frac{1}{2}(3\pi^2)^{2/3} \rho^{2/3}.
\label{eq.feg_v}
\end{equation}

Third, the non-negativity ensures
\begin{equation}
F_{\rm{\theta}}(r) \geq 0 
\end{equation}
and 
\begin{equation}
V_{\rm{\theta}}(r) \geq 0. 
\end{equation}

In order to train the MPN KEDF, we collect the Pauli energy and Pauli potential data from KSDFT calculations performed on a set of selected systems. In detail,
with the help of the Kohn-Sham orbitals and eigenvalues, in a spin degenerate system, the Pauli energy density can be analytically expressed by~\cite{88PRA-Levy-pauli}
\begin{equation}
    \tau_{\theta}^{\rm{KS}} = \sum_{i=1}^M {f_i|\nabla \psi_i(r)|^2} - \frac{|\nabla\rho|^2}{8\rho},
    \label{eq.pauli_e}
\end{equation}
while the Pauli potential has the form of
\begin{equation}
    V_{\theta}^{\rm{KS}} = \rho^{-1} \left( \tau_{\theta}^{\rm{KS}} +2 \sum_{i=1}^M {f_i(\varepsilon_M-\varepsilon_i)\psi_i^*\psi_i}\right),
    \label{eq.pauli_p}
\end{equation}
where $\psi_i(r)$ denotes an occupied Kohn-Sham orbital with index $i$, while $\varepsilon_i$ and $f_i$ are the corresponding eigenvalue and occupied number, respectively. In addition, $M$ represents the highest occupied state, and $\varepsilon_M$ is the eigenvalue of $\psi_M(r)$, i.e., the chemical potential.

\begin{figure*}[htbp]
    \centering

    \begin{subfigure}{0.3\textwidth}
    \centering
    \includegraphics[width=0.95\linewidth]{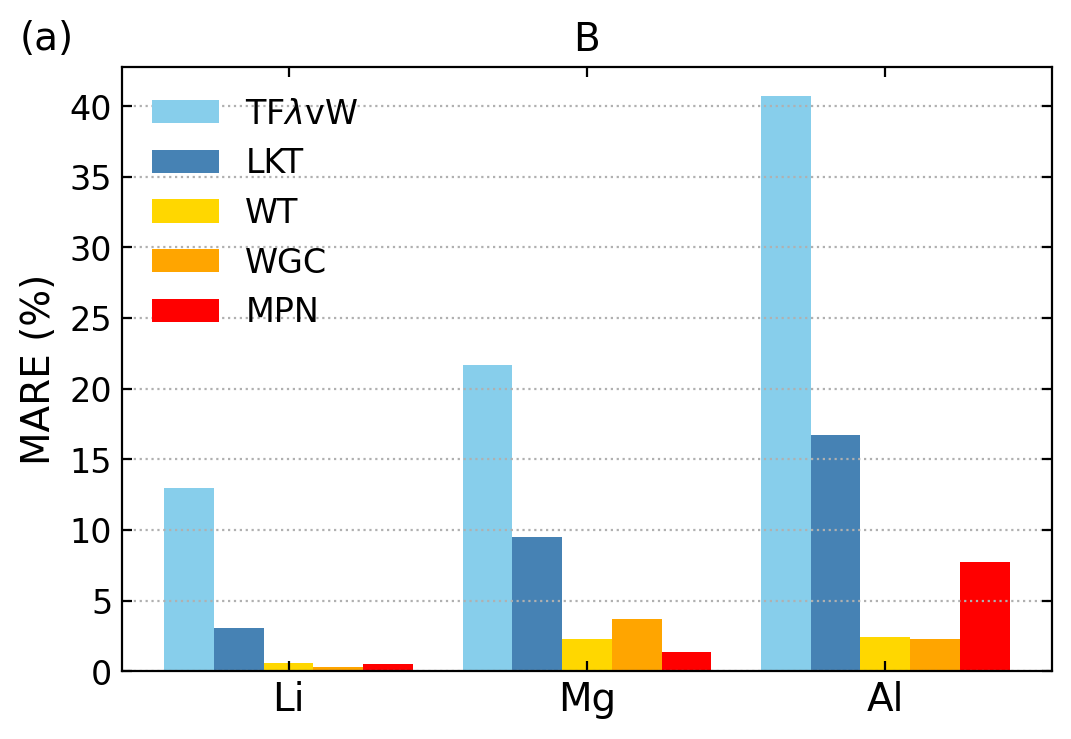}
    \label{fig:B}
    \end{subfigure}
    \begin{subfigure}{0.3\textwidth}
    \centering
    \includegraphics[width=0.95\linewidth]{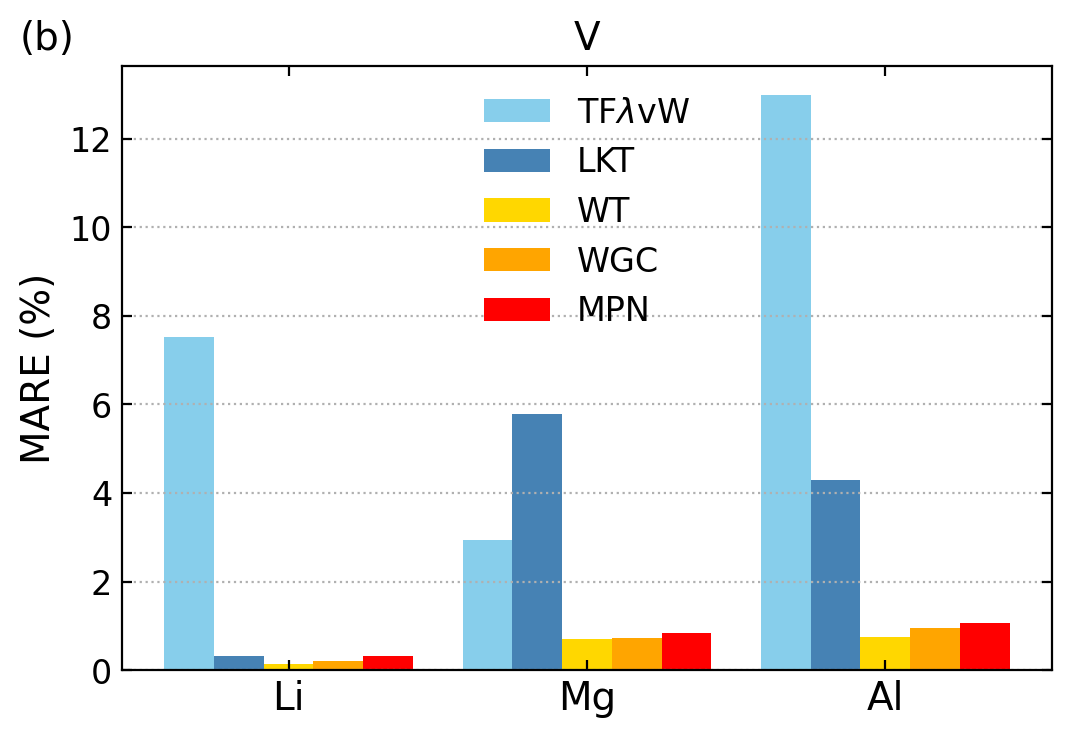}
    \label{fig:V}
    \end{subfigure}
    \begin{subfigure}{0.3\textwidth}
    \centering
    \includegraphics[width=0.95\linewidth]{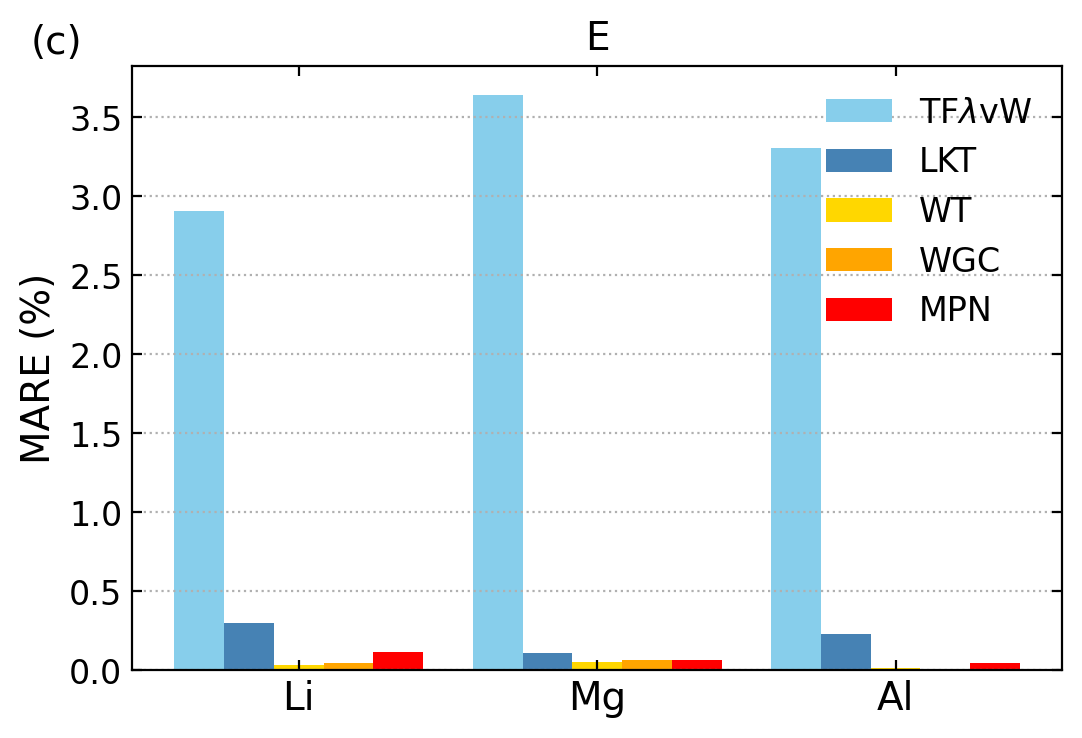}
    \label{fig:E}
    \end{subfigure}

    \caption{MAREs of bulk properties of Li, Mg, and Al systems, i.e., (a) the bulk moduli ($B$ in $\tx{GPa}$), (b) the equilibrium volumes ($V_0$ in $\tx{\AA}^3$/atom), and (c) the total energies of given systems ($E_0$ in eV/atom). 
The MARE defined in Eq.~\ref{eq.mare} is obtained by comparing OFDFT to KS-BLPS results.
We use body-centered cubic (bcc), fcc, simple cubic (sc), and cubic diamond (CD) structures of Li. We also adopt hexagonal close-packed (hcp), fcc, bcc, and sc structures of Mg. For Al systems, we take fcc, hcp, bcc, and sc structures.
    }
    \label{fig:Bulk_prop}
\end{figure*}

\begin{figure*}[htbp]
    \centering
    
    \begin{subfigure}{0.49\textwidth}
    \centering
    \includegraphics[width=0.95\linewidth]{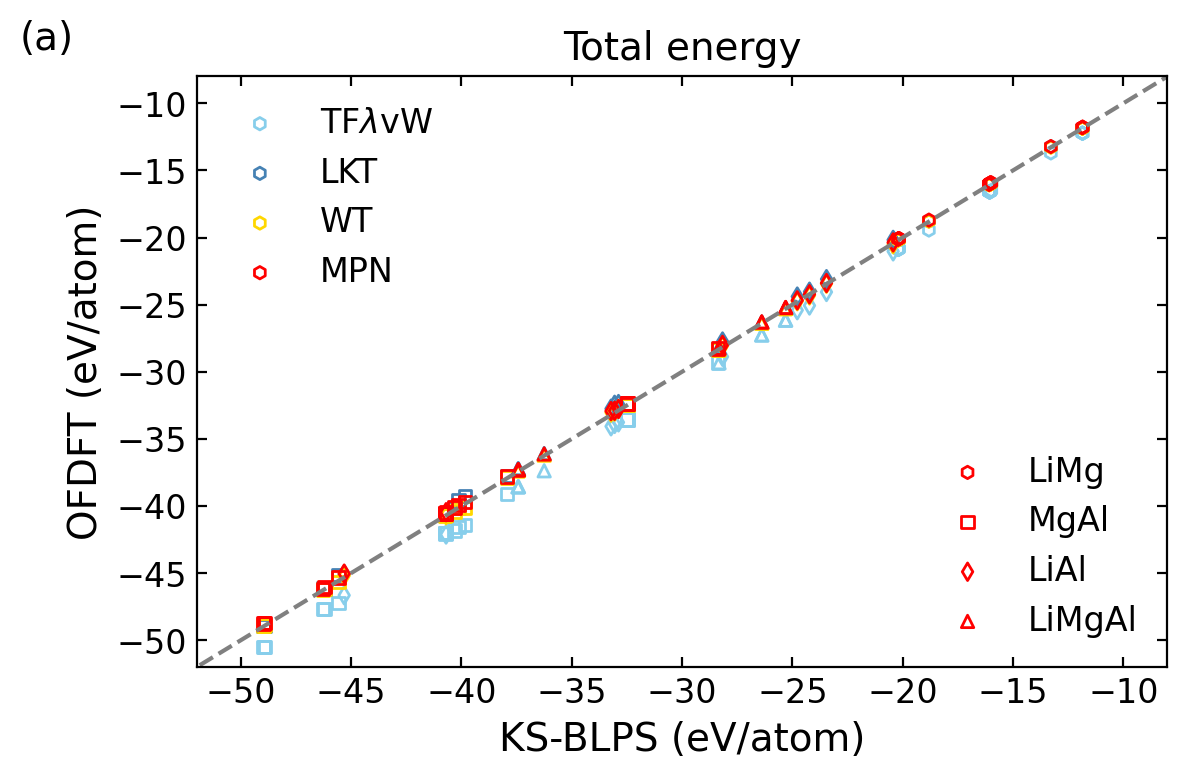}
    \label{fig:etotal}
    \end{subfigure}
    \begin{subfigure}{0.49\textwidth}
    \centering
    \includegraphics[width=0.95\linewidth]{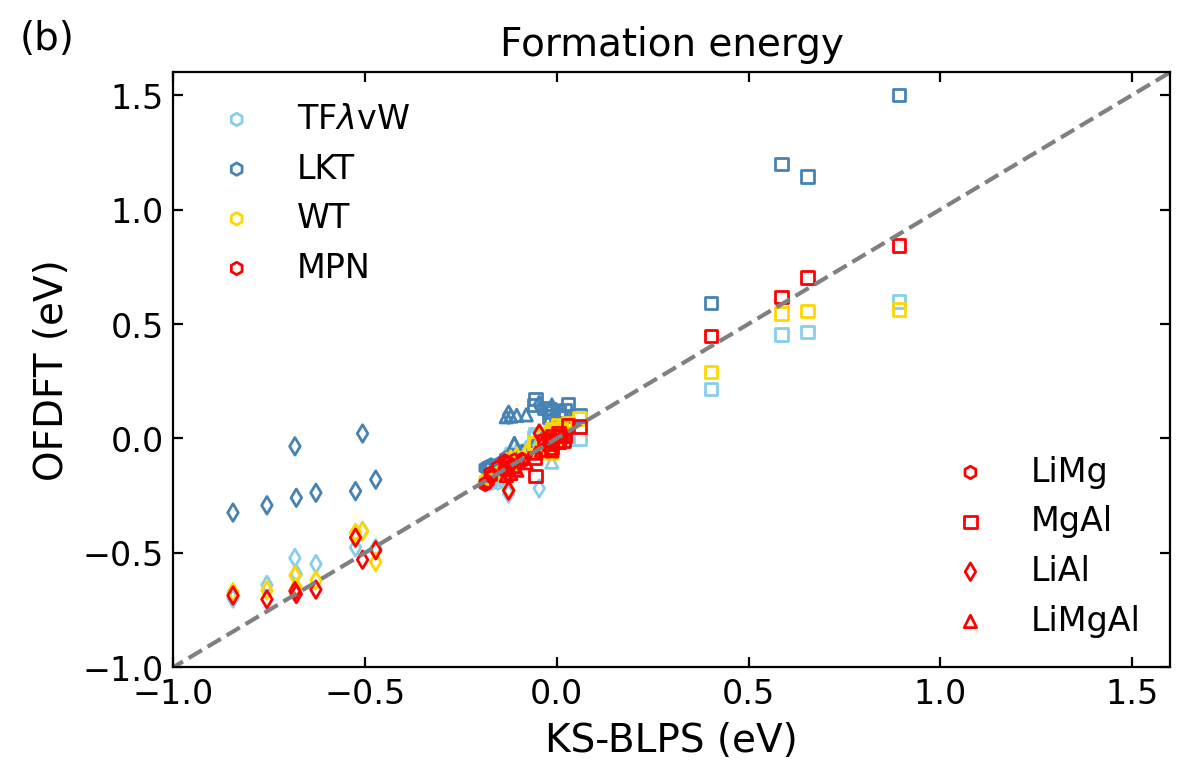}
    \label{fig:eform}
    \end{subfigure}
    
    \caption{\MC{(a) Total energies (in eV/atom) and (b) formation energies (in eV) of 59 alloys, including 20 Li-Mg alloys, 20 Mg-Li alloys, 10 Li-Al alloys, and 9 Li-Mg-Al alloys.
    Different colors indicate the formation energies from different KEDFs (TF$\lambda$vW, LKT, WT, and MPN), while different shapes of markers indicate different alloys.}}
    \label{fig:Alloys}
\end{figure*}

\subsection{Design Neural Network based on Exact Physical Constraints}

\MC{
The workflow of the MPN KEDF is summarized in Fig.~\ref{fig:workflow}.} The major structure of the MPN KEDF is an NN composed of one input layer consisting of four nodes, three hidden layers with ten nodes in each layer, and an output layer with one node.
The activation functions used in the hidden layers are chosen to be hyperbolic tangent functions, i.e., $\tanh(x)$.
In order to ensure that the calculated Pauli energy and potential obey the physical constraints mentioned above, the output of the NN is chosen as the enhancement factor $F_{\rm{\theta}}$ for each real-space grid point $r$, which is denoted as $F^{\rm{NN}}(r)$.
%
%
Next, we elucidate how nonlocal information and exact physical constraints can be incorporated into the NN to improve its accuracy and reliability.

As shown in Fig.\ref{fig:workflow}, we define four descriptors $\{\Tilde{p}, \Tilde{p}_{\rm{nl}}, \Tilde{\xi}, \Tilde{\xi}_{\rm{nl}}\}$ ({\it vide infra}) as the input of the NN for the MPN KEDF. The first descriptor $\Tilde{p}(r)$ is semilocal, while the other three are nonlocal.
First, the semilocal descriptor is the normalized dimensionless gradient of the charge density given by
\begin{equation}
    \Tilde{p}(r) = \tanh{\Big(\chi_p p(r)\Big)}, 
\end{equation}
where the parameter $p(r)$ is evaluated via
\begin{equation}
    p(r) = |\nabla \rho(r)|^2 / \Big[2(3\pi^2)^{1/3} \rho^{4/3}(r)\Big]^2.
\end{equation}
Here, $\chi_p$ is a hyper-parameter to control the distribution of $\Tilde{p}$.

Second, we propose a nonlocal descriptor of $\Tilde{p}$, which is defined as
\begin{equation}
    \Tilde{p}_{\rm{nl}}(r) = \int{w(r-r')\Tilde{p}(r'){\rm{d}^3} r'},
\end{equation}
where $w(r-r')$ is the kernel function similar to the Wang-Teter~\cite{92B-Wang-nonlocal} kernel function, satisfying 
\begin{equation}
\int{w(r-r'){\rm{d}}^3 r'}=0.
\label{eq:zero}
\end{equation}
The kernel function is defined in reciprocal space as
\begin{equation}
    w(\eta) = {{\left( {\frac{1}{2} + \frac{1-\eta^2}{4\eta}\ln \left| {\frac{1 + \eta}{1 - \eta}} \right|} \right)}^{ - 1}} - 3\eta^2 - 1.
    \label{eq.kernel}
\end{equation}
Here $\eta = \frac{k}{2k_{\tx{F}}}$ is a dimensionless reciprocal space vector, while $k_{\tx{F}} = (3\pi^2\rho_0)^{1/3}$ is the Fermi wave vector with $\rho_0$ being the average charge density.

The third and fourth nonlocal descriptors represent the distribution of charge density and take the form of
\begin{equation}
    \Tilde{\xi}(r) = \tanh{\left(\frac{\int{w(r-r')\rho^{1/3}(r'){\rm{d}^3} r'}}{\rho^{1/3}(r)}\right)},\\
\end{equation}
and
\begin{equation}
    \Tilde{\xi}_{\rm{nl}}(r) = \int{w(r-r')\Tilde{\xi}(r'){\rm{d}^3} r'},
\end{equation}
respectively.

In summary, the MPN KEDF is characterized by the above four descriptors: $\{\Tilde{p}, \Tilde{p}_{\rm{nl}}, \Tilde{\xi}, \Tilde{\xi}_{\rm{nl}}\}$, with $\chi_p=0.2$ \MC{being an empirical parameter} adopted in all calculations.
Next, we propose three physical constraints that are met by our ML-based MPN KEDF.

First, the scaling law of non-interacting electron kinetic energy is ensured when we design the above descriptors.
In detail, under the scaling translation $\rho(r) \rightarrow \rho_{\lambda}=\lambda^3\rho(\lambda r)$, the descriptors $\{\Tilde{p}(r), \Tilde{p}_{\rm{nl}}(r), \Tilde{\xi}(r), \Tilde{\xi}_{\rm{nl}}(r)\}$ become $\{\Tilde{p}(\lambda r), \Tilde{p}_{\rm{nl}}(\lambda r), \Tilde{\xi}(\lambda r), \Tilde{\xi}_{\rm{nl}}(\lambda r)\}$, i.e., the descriptors are invariant under the scaling transformation, and the detailed derivation can be found in the Supplemental Material (SM)~\cite{SM}.
Since the $T_{\rm{vW}}$ term satisfies the scaling law, we have
\begin{equation}
    \begin{aligned}
    T_{\rm{MPN}}[\rho_\lambda]
    =& T_{\rm{vW}}[\rho_\lambda] + \lambda^5 \int{\tau_{\rm{TF}}(\lambda r)}\\
    &F_{\rm{\theta}}^{\rm{NN}} \left(\Tilde{p}(\lambda r), \Tilde{p}_{\rm{nl}}(\lambda r), \Tilde{\xi}(\lambda r), \Tilde{\xi}_{\rm{nl}}(\lambda r)\right) {\rm{d}}^3{r}\\
    =& \lambda^2 \bigg[ T_{\rm{vW}}[\rho] + \int{\tau_{\rm{TF}}(\lambda r)}\\
    &F_{\rm{\theta}}^{\rm{NN}}\left(\Tilde{p}(\lambda r), \Tilde{p}_{\rm{nl}}(\lambda r), \Tilde{\xi}(\lambda r), \Tilde{\xi}_{\rm{nl}}(\lambda r)\right){\rm{d}}^3 (\lambda r) \bigg]\\
    =& \lambda^2 T_{\rm{MPN}}[\rho].
    \end{aligned}
\end{equation}
In order to verify the scaling law, we obtain the ground-state charge density $\rho(r)$ of fcc Al with the MPN KEDF, then the kinetic energy of $\rho_{\lambda}=\lambda^3\rho(\lambda r)$ with various $\lambda$ (0.25, 0.5, 1.0, 2.0, and 3.0) are calculated by the MPN KEDF.
As displayed in Fig.~\ref{fig:Scaling}, all of the $T_{\rm{MPN}}[\rho_\lambda]$s computed by the MPN KEDF fall on the line of $f(\lambda) = \lambda^2 T_s[\rho]$, demonstrating that the MPN KEDF obeys the scaling law.

The second and third constraints, i.e., \MC{the FEG limit and the non-negativity of Pauli energy density,} are introduced through post-processing of the deep neural network.
In the FEG limit, all four descriptors become zero, and hence we define the output of the NN in this limit as $F^{\rm{NN}}|_{\rm{FEG}}$.
In addition, the enhancement factor of Pauli energy is defined as 
\begin{equation}
    F_{\rm{\theta}}^{\rm{NN}} = {\rm{softplus}}\left(F^{\rm{NN}} - F^{\rm{NN}}|_{\rm{FEG}} + \ln{(e-1)}\right),
\end{equation}
where $F^{\rm{NN}}$ is the output of NN, and 
\begin{equation}
{\rm{softplus}}(x)=\ln(1+e^x)
\end{equation}
is an activation function commonly used in machine learning, satisfying 
\begin{equation}
{\rm{softplus}}(x)\geq0
\end{equation}
and
\begin{equation}
{\rm{softplus}}(x)|_{x=\ln(e-1)}=1.
\end{equation}
By construction, the non-negativity constraint is satisfied as
\begin{equation}
F_{\rm{\theta}}^{\rm{NN}} \geq 0, 
\end{equation}
and \MC{in the FEG limit where the charge density is a constant}, we have 
\begin{equation}
    \begin{aligned}    
    F_{\rm{\theta}}^{\rm{NN}}|_{\rm{FEG}} &= {\rm{softplus}}\left(F^{\rm{NN}}|_{\rm{FEG}} - F^{\rm{NN}}|_{\rm{FEG}} + \ln{(e-1)}\right) 
    \\
    &= 1,
    \end{aligned}
\end{equation}
thereby ensuring that the FEG limit is also exactly satisfied.
We note that the selection of kernel function and descriptors guarantees that once the FEG limit of Pauli energy is met, the FEG limit of Pauli potential is automatically satisfied, as discussed in Section III of SM~\cite{SM}.

\MC{
Fig.~\ref{fig:workflow} summarizes the workflow of the MPN KEDF, which involves the abovementioned physical constraints.
First, for each real-space grid point, the descriptors of charge density $\rho(r)$ ($\{\Tilde{p}, \Tilde{p}_{\rm{nl}}, \Tilde{\xi}, \Tilde{\xi}_{\rm{nl}}\}$) are entered into NN to get the corresponding enhancement factor $F^{\rm{NN}}(r)$. 
Second, the descriptors of FEG ($\{\Tilde{p}=0, \Tilde{p}_{\rm{nl}}=0, \Tilde{\xi}=0, \Tilde{\xi}_{\rm{nl}}=0\}$) are fed into the NN, and the enhancement factor of FEG $F^{\rm{NN}}|_{\rm{FEG}}$ is obtained.
Third, to ensure both the FEG limit and the non-negativity of Pauli energy density are satisfied, the enhancement factor of Pauli energy is defined as $F_{\rm{\theta}}^{\rm{NN}} = {\rm{softplus}}\left(F^{\rm{NN}} - F^{\rm{NN}}|_{\rm{FEG}} + \ln{(e-1)}\right)$. 
Finally, the kinetic energy and kinetic potential are calculated by the MPN KEDF using the defined formulas.
}

\begin{table*}[htbp]
	\centering
	\caption{MAEs (Eq.~\ref{eq.mae}) of the total energies and formation energies of 59 alloys obtained by \MC{comparing various KEDFs (TF$\lambda$vW, LKT, WT, and MPN) in OFDFT} to KS-BLPS results. The systems include 20 Li-Mg alloys, 20 Mg-Li alloys, 10 Li-Al alloys, and 9 Li-Mg-Al alloys.}
	\begin{tabularx}{0.9\linewidth}{
			>{\raggedright\arraybackslash\hsize=2\hsize\linewidth=\hsize}X
			>{\centering\arraybackslash\hsize=0.8\hsize\linewidth=\hsize}X
			>{\centering\arraybackslash\hsize=0.8\hsize\linewidth=\hsize}X
			>{\centering\arraybackslash\hsize=0.8\hsize\linewidth=\hsize}X
			>{\centering\arraybackslash\hsize=0.8\hsize\linewidth=\hsize}X
			>{\centering\arraybackslash\hsize=0.8\hsize\linewidth=\hsize}X}
		\hline\hline
		MAE of total energy (eV/atom)       &LiMg    &MgAl    &LiAl    &LiMgAl &Total\\
		\hline
		TF$\rm{\lambda}$vW &0.540  &1.330  &0.873  &0.999  &0.934\\
            LKT               &0.040  &0.156  &0.351  &0.124  &0.145\\
            WT                &0.013  &0.059  &0.082  &0.031  &0.043\\
            MPN                &0.078  &0.163  &0.146  &0.106  &0.123\\
            \hline
		MAE of formation energy (eV)       &LiMg    &MgAl    &LiAl    &LiMgAl &Total\\
		\hline
		TF$\rm{\lambda}$vW &0.022  &0.061  &0.103  &0.036  &0.051\\
            LKT               &0.041  &0.189  &0.397  &0.138  &0.166\\
            WT                &0.005  &0.050  &0.077  &0.023  &0.035\\
            MPN                &0.015  &0.027  &0.056  &0.024  &0.028\\
		\hline\hline
	\end{tabularx}
	\label{tab:Alloy_e}
\end{table*}

\begin{table*}[htbp]
	\centering
	\caption{Mean MAREs (Eq.~\ref{eq.mare}) of charge densities of 59 alloys, including 20 Li-Mg alloys, 20 Mg-Li alloys, 10 Li-Al alloys, and 9 Li-Mg-Al alloys. MAREs are obtained by comparing \MC{various KEDFs (TF$\lambda$vW, LKT, WT, and MPN) in} OFDFT to KS-BLPS results.
        }
	\begin{tabularx}{0.9\linewidth}{
			>{\raggedright\arraybackslash\hsize=2\hsize\linewidth=\hsize}X
			>{\centering\arraybackslash\hsize=0.8\hsize\linewidth=\hsize}X
			>{\centering\arraybackslash\hsize=0.8\hsize\linewidth=\hsize}X
			>{\centering\arraybackslash\hsize=0.8\hsize\linewidth=\hsize}X
			>{\centering\arraybackslash\hsize=0.8\hsize\linewidth=\hsize}X
			>{\centering\arraybackslash\hsize=0.8\hsize\linewidth=\hsize}X}
		\hline\hline
		mean MARE of charge density ($\%$)       &LiMg    &MgAl    &LiAl    &LiMgAl &Total\\
		\hline
		TF$\rm{\lambda}$vW    &12.40 &16.11 &13.26 &15.66 &14.30\\
            LKT                   &5.26  &7.44  &11.61 &6.98  &7.34\\
            WT                    &1.06  &2.57  &4.98  &2.04  &2.38\\
            MPN                    &2.41  &3.12  &5.81  &2.89  &3.30\\
		\hline\hline
	\end{tabularx}
	\label{tab:Density}
\end{table*}

\begin{figure*}[thbp]
    \centering

    \begin{subfigure}{0.49\textwidth}
    \centering
    \includegraphics[width=0.95\linewidth]{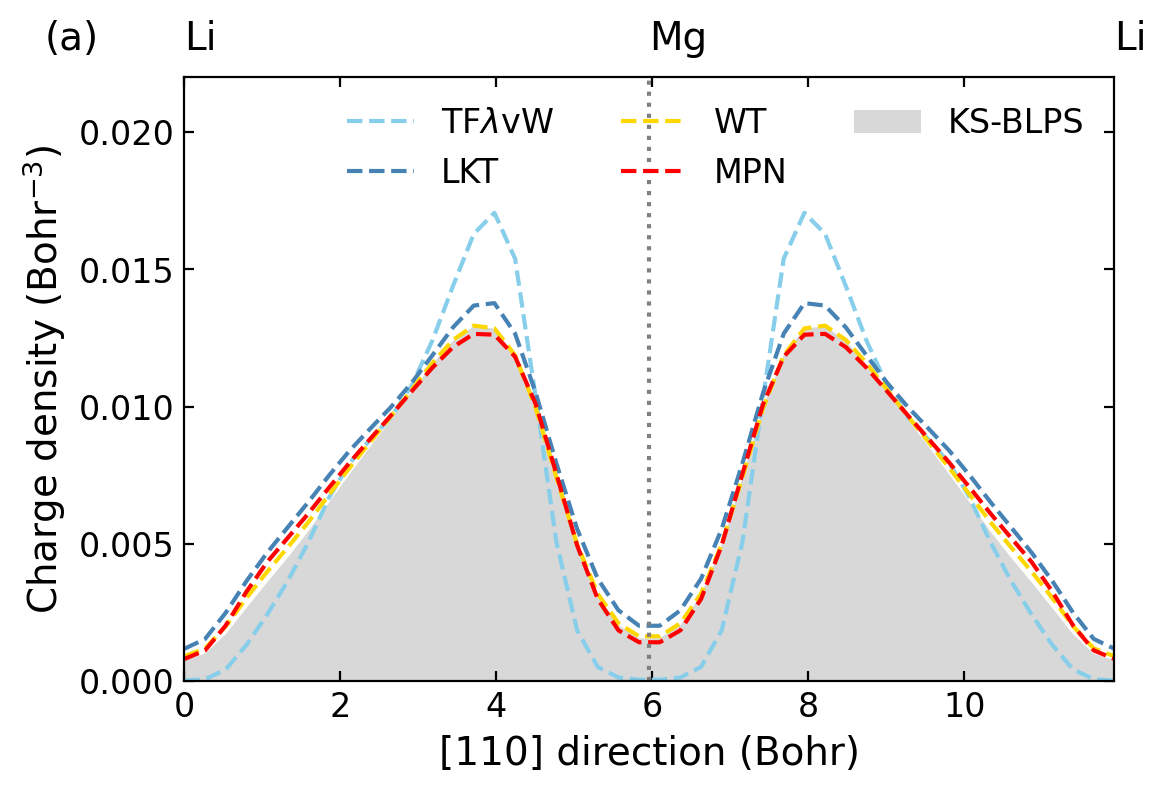}
    \label{fig:mp-976254}
    \end{subfigure}
    \begin{subfigure}{0.49\textwidth}
    \centering
    \includegraphics[width=0.95\linewidth]{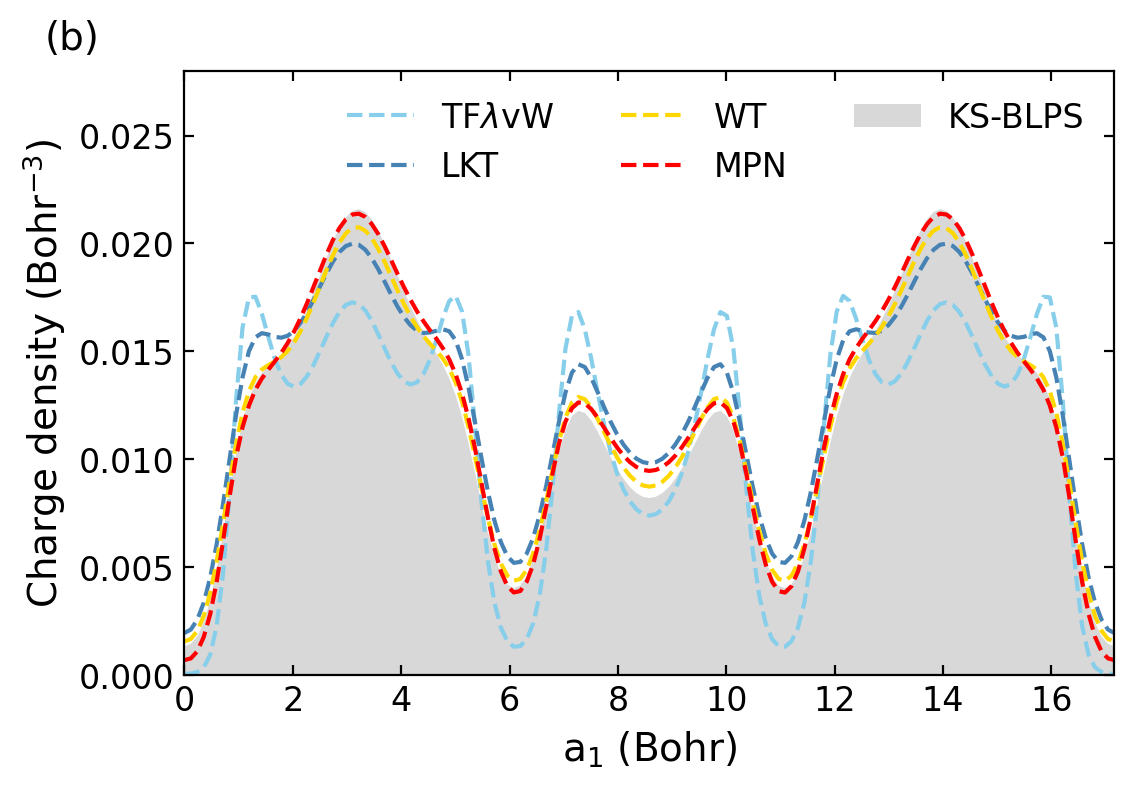}
    \label{fig:mp-1185175}
    \end{subfigure}
    \begin{subfigure}{0.49\textwidth}
    \centering
    \includegraphics[width=0.95\linewidth]{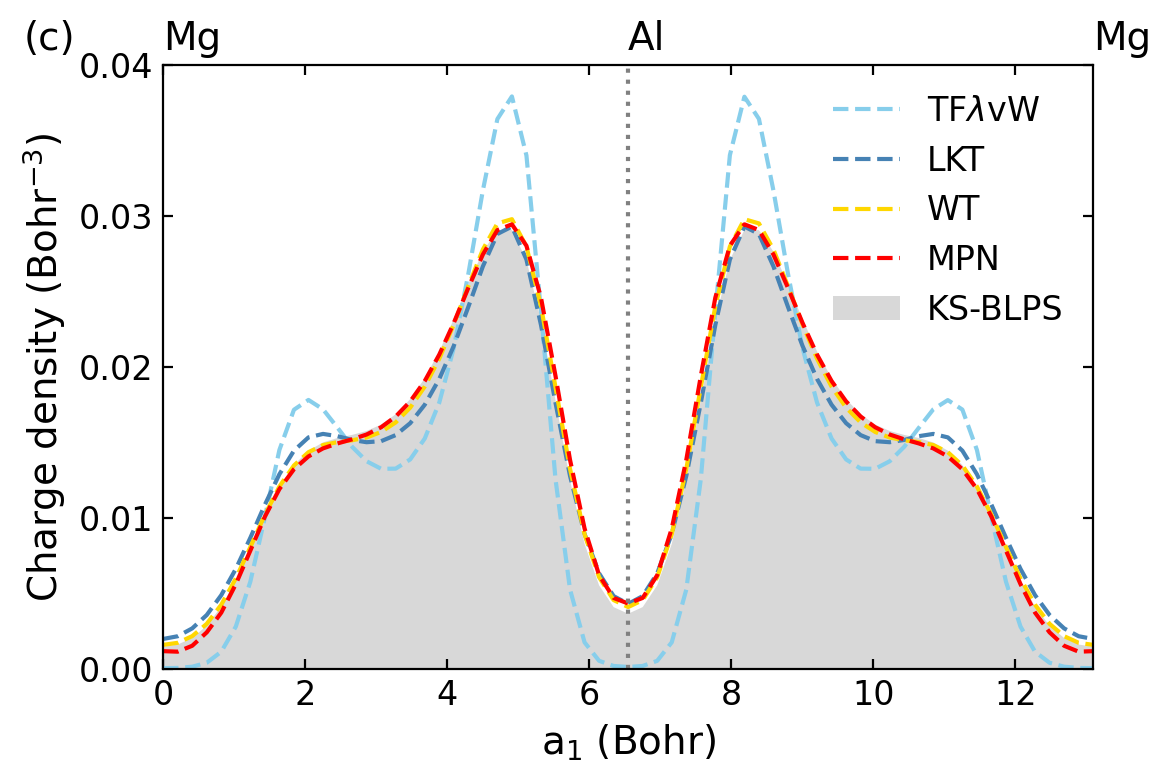}
    \label{fig:mp-1094666}
    \end{subfigure}
    \begin{subfigure}{0.49\textwidth}
    \centering
    \includegraphics[width=0.95\linewidth]{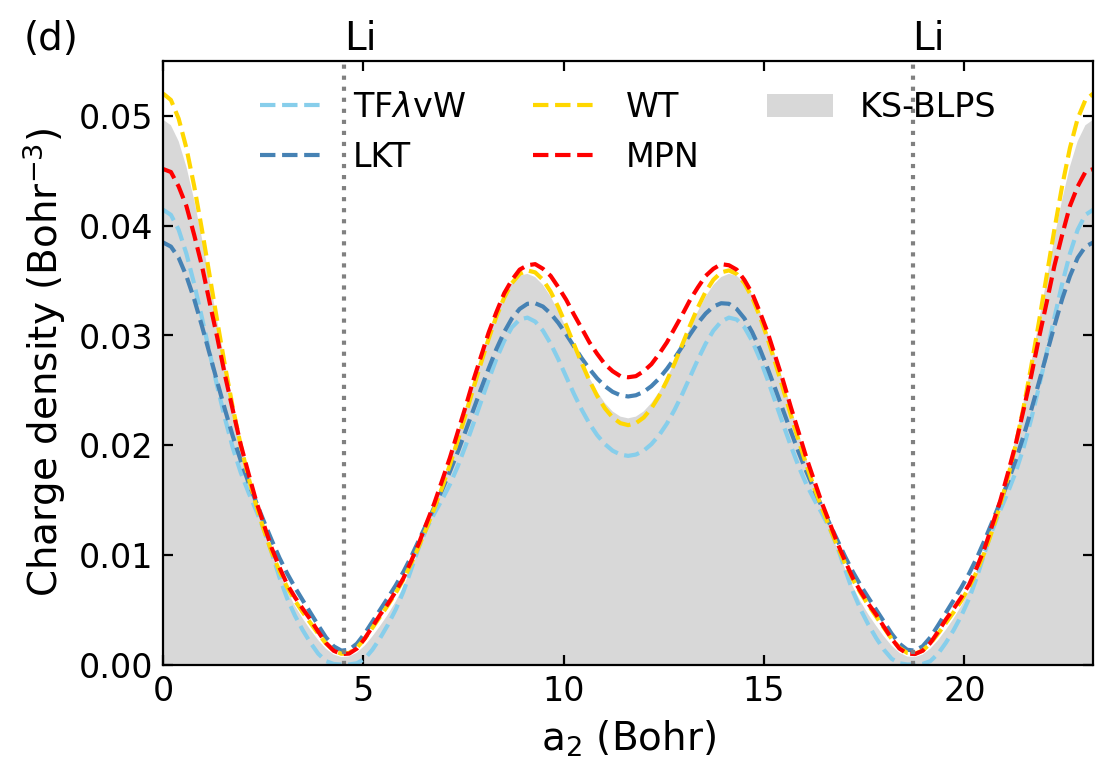}
    \label{fig:mp-1191737}
    \end{subfigure}

    \caption{Charge densities of four typical alloys.
    (a) $\rm{Li_3 Mg}$ (mp-976254, 4 atoms) from the training set. The MARE of charge density obtained by the TF$\rm{\lambda}$vW, LKT, WT, and MPN KEDFs are 13.85\%, 5.28\%, 1.03\%, and 2.73\%, respectively.
    (b) $\rm{Li(Mg_4 Al_3)_4}$ (mp-1185175), the largest system in the testing set, containing 87 atoms. The MARE of charge density obtained by the TF$\rm{\lambda}$vW, LKT, WT, and MPN KEDFs are 16.15\%, 7.76\%, 2.54\%, and 3.08\%, respectively.
    (c) $\rm{Mg_3 Al}$ (mp-1094666, 16 atoms) from the testing set, in which the MPN KEDF yields the lowest MARE among the testing set. The MARE of charge density obtained by the TF$\rm{\lambda}$vW, LKT, WT, and MPN KEDFs are 16.73\%, 6.42\%, 1.65\%, and 1.57\%, respectively.
    (d) $\rm{LiAl}$ (mp-1191737, 48 atoms) from the testing set, in which the MPN KEDF obtains the largest MARE among the testing set. The MARE of charge density obtained by the TF$\rm{\lambda}$vW, LKT, WT, and MPN KEDFs are 13.51\%, 15.41\%, 6.98\%, and 8.16\%, respectively.
    The labels $a_1$ and $a_2$ denote the first and second lattice vectors of the corresponding structures, respectively.
    }
    \label{fig:Density}
\end{figure*}

\subsection{Training Details}

\MC{Before training the MPN KEDF,} the loss function is defined as
\begin{equation}
    \begin{aligned}
        L=&\frac{1}{N}\sum_r{\left[ \left(\frac{F_\theta^{\rm{NN}}- F^{\rm{KS}}_{\theta}}{\bar{F}^{\rm{KS}}_{\theta}}\right)^2 +
      \left(\frac{V_\theta^{\rm{MPN}} - V^{\rm{KS}}_{\theta}}{\bar{V}^{\rm{KS}}_{\theta}}\right)^2 \right]}\\
      &+ \left[F^{\rm{NN}}|_{\rm{FEG}}-\ln(e-1)\right]^2.
    \end{aligned}
\end{equation}
Where $N$ is the number of grid points, and $\bar{F}^{\rm{KS}}_{\theta}$ ($\bar{V}^{\rm{KS}}_{\theta}$) represents the mean of $F^{\rm{KS}}_{\theta}$ ($V^{\rm{KS}}_{\theta}$).
The first term helps NN to learn information from the Pauli energy, while the second term emphasizes the significance of reproducing the correct Pauli potential.
We emphasize that fitting the Pauli potential is crucial in determining the optimization direction and step length during the OFDFT calculations, and $V_\theta^{\rm{MPN}}$ can be obtained through the back propagation of NN, as derived in SM~\cite{SM}.
The last term is a penalty term to reduce the magnitude of the FEG correction, which improves the stability of the MPN KEDF.

The training set consists of eight metallic structures, namely bcc Li, fcc Mg, fcc Al, as well as five alloys: $\rm{Li_3 Mg}$ (mp-976254), LiMg (mp-1094889), $\rm{Mg_3 Al}$ (mp-978271), $\beta''$ $\rm{MgAl_3}$~\cite{03MSMSE-Carling-mgal}, $\rm{LiAl_3}$ (mp-10890), where the numbers in brackets are the Materials Project IDs~\cite{13APL-Jain-MP}.
We performed KSDFT calculations to obtain the ground charge density and calculate the corresponding descriptors.
Additionally, the Pauli energy and potential are calculated using Eqs.~\ref{eq.pauli_e} and \ref{eq.pauli_p}, respectively.
These calculations are performed on a $27\times27\times27$ grid, resulting in a total of 157,464 grid points in the training set.

\subsection{Numerical Details}

We have employed the ABACUS 3.0.4 packages~\cite{16Li-CMS-ABACUS} to carry out OFDFT and KSDFT calculations, while for OFDFT with the Wang-Govind-Carter (WGC) KEDF~\cite{99B-Wang-nonlocal}, we have utilized the PROFESS 3.0 package.~\cite{15CPC-Chen-PROFESS}
The MPN KEDF is implemented in ABACUS using the libtorch package,~\cite{19ANIPS-paszke-pytorch}, and the libnpy package is adopted to dump the data.
\MC{Table S1} lists the plane-wave energy cutoffs employed in both OFDFT and KSDFT calculations, as well as the Monkhorst-Pack $k$-point samplings~\cite{76B-Monkhorst} utilized in KSDFT.
For both OFDFT and KSDFT calculations, we used the Perdew-Burke-Ernzerhof (PBE)~\cite{96PRL-Perdew-PBE} and bulk-derived local pseudopotentials (BLPS)~\cite{08PCCP-Huang-BLPS}.
Additionally, we used the Gaussian smearing method with a smearing width of 0.1 eV in our KSDFT calculations.

In order to calculate the ground-state bulk properties, we first optimize the crystal structures until the stress tensor elements are below $5\times10^{-7}$\ Hartree/$\text{Bohr}^3$, then compress and expand the \MC{lattice constant of the} unit cell from $0.99a_0$ to $1.01a_0$, where $a_0$ is the equilibrium lattice constant.
Once the energy-volume curve is obtained, the bulk modulus $B$ of a given system is fitted by Murnaghan’s equation of state.\cite{44-Murnaghan-bulkmodulus}

We compare the results obtained by the MPN KEDF to those obtained from OFDFT calculations with traditional KEDFs.
Specifically, we have employed semilocal KEDFs such as the TF$\rm{\lambda}$vW~\cite{83pra-berk-semilocal} and the Luo-Karasiev-Trickey (LKT) KEDFs~\cite{18B-Luo-semilocal}, as well as the nonlocal ones including the Wang-Teter (WT)~\cite{92B-Wang-nonlocal} and WGC KEDFs. 
The parameter $\rm{\lambda}$ of TF$\rm{\lambda}$vW has been set to be 0.2, and the parameter $a$ of the LKT KEDF is set to be 1.3, as in the original work~\cite{18B-Luo-semilocal}.
In addition, we set $\alpha$=$\frac{5+\sqrt{5}}{6}$, $\beta$=$\frac{5-\sqrt{5}}{6}$ and $\gamma$=2.7 in the WGC KEDF~\cite{99B-Wang-nonlocal}, as well as $\alpha$=$\frac{5}{6}$, $\beta$=$\frac{5}{6}$ in the WT KEDF~\cite{92B-Wang-nonlocal}.
%

The formation energy $E_{\rm{f}}$ of Li-Mg-Al alloy is defined as
\begin{equation}
    E_{\rm{f}}=\frac{1}{N}\left(E_{\rm{total}} - n_{\rm{Li}} E_{\rm{Li}} - n_{\rm{Mg}} E_{\rm{Mg}} - n_{\rm{Al}} E_{\rm{Al}}\right),
\label{eq.eform}
\end{equation}
\MC{where $E_{\rm{total}}$ is the total energy of the alloy, and $E_{\rm{Li}}$, $E_{\rm{Mg}}$, and $E_{\rm{Al}}$ denote the equilibrium energy of the bcc Li, hcp Mg, and fcc Al structures, respectively.} 
Furthermore, $n_{\rm{Li}}$, $n_{\rm{Mg}}$, and $n_{\rm{Al}}$ depict the number of Li, Mg, and Al atoms, respectively. $N=n_{\rm{Li}}+n_{\rm{Mg}}+n_{\rm{Al}}$ \MC{denotes} the total number of atoms of the alloy.

The Mean Absolute Relative Error (MARE) and Mean Absolute Error (MAE) of property $x$ are respectively defined as 
\begin{equation}
    {\rm{MARE}}=\frac{1}{N}\sum_i^N{\left|\frac{x_i^{\rm{OF}} - x_i^{\rm{KS}}}{x_i^{\rm{KS}}}\right|},\\
    \label{eq.mare}
\end{equation}
\begin{equation}
    {\rm{MAE}}=\frac{1}{N}\sum_i^N{\left|x_i^{\rm{OF}} - x_i^{\rm{KS}}\right|}.
    \label{eq.mae}
\end{equation}
Here $N$ is the number of data points, $x_i^{\rm{OF}}$ and $x_i^{\rm{KS}}$ are \MC{properties} obtained from OFDFT and KSDFT \MC{calculations}, respectively.

\section{Results and Discussion}

In order to examine the precision and transferability of the MPN KEDF, we prepared two testing sets.
The first set comprises 4 structures of Li (bcc, fcc, sc, and CD), 4 structures of Mg (hcp, fcc, bcc, and sc), and 4 structures of Al (fcc, hcp, bcc, and sc). We evaluated the properties of these bulk systems, including the bulk moduli, the equilibrium volumes, and the equilibrium energies using various KEDFs.
For the second testing set, we selected 59 alloys obtained from the Materials Project database~\cite{13APL-Jain-MP}, including 20 Li-Mg alloys, 20 Mg-Li alloys, 10 Li-Al alloys, and 9 Li-Mg-Al alloys, and the detailed information of these alloys are listed in Table S2.

Notably, \MC{most of the structures in the two testing sets do not appear in the training set}, allowing for an unbiased comparison.
We systematically compared the total energies, the formation energies, and the charge densities of alloys within the \MC{second testing set.}
We also trained \MC{another} semilocal ML-based KEDF with descriptors as $\{\Tilde{p}, \Tilde{q}\}$ with $\Tilde{q} = \tanh{(0.1 q)}$, \MC{where} $q=\nabla^2 \rho / [4(3\pi^2)^{2/3} \rho^{5/3}]$.
However, \MC{we found} this semilocal ML-based KEDF cannot achieve convergence in all tested systems.

\subsection{Simple Metals}

Fig.~\ref{fig:Bulk_prop} displays the MAREs of bulk properties of Li, Mg, and Al systems.
Compared to the nonlocal WT and WGC KEDFs, the semilocal KEDFs \MC{(the TF$\rm{\lambda}$vW and LKT KEDFs)} yield larger MAREs across all the properties in all three systems, indicating that the nonlocal information is crucial to enhance the accuracy of KEDF.
Comparatively, the MPN KEDF yields MAREs slightly larger than those of the WT and WGC KEDFs but does not exceed those of semilocal ones.
Notably, the  MPN KEDF achieves a lower MARE of 1.37\% for the bulk modulus of Mg, outperforming WT and WGC KEDFs, which exhibit MAREs of 2.32\% and 3.73\%, respectively.
On the other hand, the poorest results obtained by the MPN KEDF are the bulk modulus of Al, where it exhibits a MARE of 7.75\%, nearly three times than those from the WT (2.41\%) and WGC (2.26\%) KEDFs. \MC{This may be caused by the fact that we did not include more Al structures with different densities in the training set.}
However, even in this case, the MAREs obtained by the TF$\rm{\lambda}$vW (40.72\%) and LKT (16.69\%) KEDFs are almost five and two times higher than that of the MPN KEDF.
As a result, we conclude that the MPN KEDF yields reasonable accuracy when compared to other nonlocal KEDFs.

It is noteworthy that the energy difference between the fcc and hcp structures of bulk Al is small, which is 0.025 eV/atom as predicted by KSDFT, and it is sensitive to the accuracy of KEDF.~\cite{23B-Sun-TKK}
Both TF$\rm{\lambda}$vW and LKT KEDFs, as semilocal KEDFs, fail to distinguish this subtle energy difference and predict it as 0.000 eV/atom.
In contrast, the nonlocal WT and WGC KEDFs yield non-zero energy differences of 0.018 and 0.016 eV/atom, respectively.
Moreover, the MPN KEDF predicts the energy difference to be 0.021 eV/atom, which is close to the result of \MC{0.025 eV/atom} obtained by KSDFT and is more accurate than the WT and WGC KEDFs.
This result emphasizes the importance of involving the nonlocal information again, which enables the MPN KEDF to distinguish the subtle difference between similar crystal structures.\cite{23B-Sun-TKK}

\subsection{Alloys}

Fig.~\ref{fig:Alloys} illustrates the total energies and the formation energies of 59 alloys obtained by different KEDFs in OFDFT calculations, and their corresponding MAEs are listed in Table.~\ref{tab:Alloy_e}.
Notably, the WGC KEDF failed to achieve convergence for nine alloys; therefore, we have excluded \MC{the WGC results from Table.~\ref{tab:Alloy_e}.}
Regarding the prediction of total energy shown in Fig.~\ref{fig:Alloys}(a), the TF$\rm{\lambda}$vW KEDF consistently underestimates the values compared to those obtained by KSDFT, resulting in a large MAE of 0.934 eV/atom.
In contrast, the LKT KEDF performs better with a reduced MAE of 0.145 eV/atom, while the nonlocal WT KEDF yields a lower MAE of 0.043 eV/atom. 
The MPN KEDF yields a higher MAE (0.123 eV/atom) than the WT KEDF but still outperforms the TF$\rm{\lambda}$vW and LKT KEDFs.
As for the formation energies shown in Fig.~\ref{fig:Alloys}(b), we observe that the LKT KEDF consistently yields larger values compared to KSDFT, resulting in a high MAE of 0.166 eV, which is much larger than the MAEs obtained by \MC{the TF$\rm{\lambda}$vW KEDF (0.051 eV) and WT KEDFs (0.035 eV). 
Remarkably}, the MPN KEDF exhibits an even lower  MAE (0.028 eV) than the WT KEDF.
Overall, these results demonstrate the promising potential of the MPN KEDF in predicting the energies of complex alloy systems with a high accuracy.

In order to further evaluate the accuracy of the MPN KEDF, we computed the charge densities of 59 alloys and calculated the mean MAREs, listed in Table~\ref{tab:Density}.
As expected, the semilocal TF$\rm{\lambda}$vW and LKT KEDFs failed to reproduce the charge density obtained by KSDFT, exhibiting mean MAREs of 14.30\% and 7.34\%, respectively. These MAREs are considerably higher than the mean MARE obtained by the nonlocal WT KEDF (2.38\%).
\MC{Impressively,} the MPN KEDF yields a mean MARE of 3.30\%, which is slightly higher than that of the WT KEDF but significantly lower than those of the TF$\rm{\lambda}$vW and LKT KEDFs.
We note that the above 59 alloys are not present in the training set, and there are even no Li-Mg-Al alloys in the training set, so the above results not only indicate a high accuracy but also excellent transferability of the MPN KEDF.

Fig.~\ref{fig:Density} shows the charge densities of four typical structures, one taken from the training set and the other three from the testing set.
The first structure is $\rm{Li_3 Mg}$ (mp-976254) from the training set, containing four atoms.
The MPN KEDF yields a MARE of 2.73\%, which is slightly larger than that obtained by the WT KEDF (1.03\%) but significantly lower than those obtained by the TF$\rm{\lambda}$vW  (13.85\%) and LKT KEDFs (5.28\%), demonstrating the efficiency of the training process.

The second structure is $\rm{Li(Mg_4 Al_3)_4}$ (mp-1185175) \MC{with 87 atoms, which is the largest system among the testing set.}
Notably, the MPN KEDF \MC{achieves convergence to yield a smooth ground-state density}, which is close to the result obtained by KSDFT, indicating \MC{an excellent stability in optimizing the electron charge density}. 
In contrast, the WGC KEDF fails to reach convergence for this structure.

The last two \MC{crystal structures} are $\rm{Mg_3 Al}$ (mp-1094666, 16 atoms) and $\rm{LiAl}$ (mp-1191737, 48 atoms) from the testing set, for which the MPN KEDF yields the lowest MARE and largest MARE among the testing set, respectively.
For the $\rm{Mg_3 Al}$ structure, the MPN KEDF \MC{exhibits a better accuracy than the WT KEDF}, yielding a MARE of 1.57\%, lower than the 1.65\% obtained by the WT KEDF. 
For the LiAl structure, although the MPN KEDF yields the largest MARE of 8.16\%, it is still much lower than those obtained by the TF$\rm{\lambda}$vW (13.51\%) and LKT KEDFs (15.41\%).
Overall, the MPN KEDF outperforms the semilocal KEDFs in terms of accuracy and achieves comparable accuracy to the other nonlocal KEDFs.
Additionally, the stability of the MPN KEDF is evidenced by reaching convergence and obtaining smooth charge densities for all alloys in the testing set.

\SL{What's more, in order to further test the transferability and stability of the MPN KEDF, we have generated 45 hypothetical Mg-Al alloys and calculated the total energies and formation energies of these alloys.
Similar to the phenomenon described above, the MPN KEDF gives worse total energies than the WT KEDF, but it yields substantially more accurate formation energies than the WT KEDF, and detailed results can be found in SM~\cite{SM}.
In conclusion, the above results imply the good transferability of the MPN KEDF for Mg-Al alloys.}

\section{Conclusions}
\MC{
In this work, based on the framework of deep neural networks, we proposed an ML-based physical-constrained nonlocal (MPN) KEDF.
%
Our proposed method relied on four descriptors, i.e., $\{\Tilde{p}, \Tilde{p}_{\rm{nl}}, \Tilde{\xi}, \Tilde{\xi}_{\rm{nl}}\}$, in which $\Tilde{p}$ was a semilocal descriptor, and the other three captured the nonlocal information of charge density.
Importantly, the MPN KEDF was subject to three crucial physical constraints, including the scaling law of Eq.~\ref{eq.scaling}, the FEG limit shown in Eq.~\ref{eq.feg_f} 
and the non-negativity of Pauli energy density.
%
The MPN KEDF was implemented in the ABACUS package.~\cite{16Li-CMS-ABACUS}}

\MC{
We systematically evaluated the performance of various KEDFs on simple metals, including bulk Li, Mg, and Al, by calculating their bulk properties, i.e., the bulk moduli, the equilibrium volumes, and the equilibrium energies.
Additionally, we tested 59 alloys consisting of 20 Li-Mg alloys, 20 Mg-Li alloys, 10 Li-Al alloys, and 9 Li-Mg-Al alloys.
%
%
%
Overall, our results demonstrated that the MPN KEDF exceeded the accuracy of semilocal KEDFs and approached the accuracy of nonlocal KEDFs for all of the tested systems.
Additionally, the proposed MPN KEDF exhibited satisfactory transferability and stability during density optimization. 
%
}

\MC{
In the future, our proposed approach sheds new light on generating KEDFs for semiconductors or molecular systems, and may also serve as a reference for developing ML-based exchange-correlation functionals.}

\acknowledgements

The work of L.S. and M.C. was supported by the National Science Foundation of China under Grand No. 12074007 and No. 12122401. The numerical simulations were performed on the High-Performance Computing Platform of CAPT.

\bibliography{ML-KEDF}

\begin{thebibliography}{47}%
\makeatletter
\providecommand \@ifxundefined [1]{%
 \@ifx{#1\undefined}
}%
\providecommand \@ifnum [1]{%
 \ifnum #1\expandafter \@firstoftwo
 \else \expandafter \@secondoftwo
 \fi
}%
\providecommand \@ifx [1]{%
 \ifx #1\expandafter \@firstoftwo
 \else \expandafter \@secondoftwo
 \fi
}%
\providecommand \natexlab [1]{#1}%
\providecommand \enquote  [1]{``#1''}%
\providecommand \bibnamefont  [1]{#1}%
\providecommand \bibfnamefont [1]{#1}%
\providecommand \citenamefont [1]{#1}%
\providecommand \href@noop [0]{\@secondoftwo}%
\providecommand \href [0]{\begingroup \@sanitize@url \@href}%
\providecommand \@href[1]{\@@startlink{#1}\@@href}%
\providecommand \@@href[1]{\endgroup#1\@@endlink}%
\providecommand \@sanitize@url [0]{\catcode `\\12\catcode `\$12\catcode
  `\&12\catcode `\#12\catcode `\^12\catcode `\_12\catcode `\%12\relax}%
\providecommand \@@startlink[1]{}%
\providecommand \@@endlink[0]{}%
\providecommand \url  [0]{\begingroup\@sanitize@url \@url }%
\providecommand \@url [1]{\endgroup\@href {#1}{\urlprefix }}%
\providecommand \urlprefix  [0]{URL }%
\providecommand \Eprint [0]{\href }%
\providecommand \doibase [0]{https://doi.org/}%
\providecommand \selectlanguage [0]{\@gobble}%
\providecommand \bibinfo  [0]{\@secondoftwo}%
\providecommand \bibfield  [0]{\@secondoftwo}%
\providecommand \translation [1]{[#1]}%
\providecommand \BibitemOpen [0]{}%
\providecommand \bibitemStop [0]{}%
\providecommand \bibitemNoStop [0]{.\EOS\space}%
\providecommand \EOS [0]{\spacefactor3000\relax}%
\providecommand \BibitemShut  [1]{\csname bibitem#1\endcsname}%
\let\auto@bib@innerbib\@empty
\bibitem [{\citenamefont {Hohenberg}\ and\ \citenamefont
  {Kohn}(1964)}]{64PR-Hohenberg}%
  \BibitemOpen
  \bibfield  {author} {\bibinfo {author} {\bibfnamefont {P.}~\bibnamefont
  {Hohenberg}}\ and\ \bibinfo {author} {\bibfnamefont {W.}~\bibnamefont
  {Kohn}},\ }\bibfield  {title} {\bibinfo {title} {Inhomogeneous electron
  gas},\ }\href@noop {} {\bibfield  {journal} {\bibinfo  {journal} {Phys.
  Rev.}\ }\textbf {\bibinfo {volume} {136}},\ \bibinfo {pages} {864B} (\bibinfo
  {year} {1964})}\BibitemShut {NoStop}%
\bibitem [{\citenamefont {Kohn}\ and\ \citenamefont {Sham}(1965)}]{65PR-Kohn}%
  \BibitemOpen
  \bibfield  {author} {\bibinfo {author} {\bibfnamefont {W.}~\bibnamefont
  {Kohn}}\ and\ \bibinfo {author} {\bibfnamefont {L.~J.}\ \bibnamefont
  {Sham}},\ }\bibfield  {title} {\bibinfo {title} {Thermal properties of the
  inhomogeneous electron gas},\ }\href@noop {} {\bibfield  {journal} {\bibinfo
  {journal} {Phys. Rev.}\ }\textbf {\bibinfo {volume} {140}},\ \bibinfo {pages}
  {1133A} (\bibinfo {year} {1965})}\BibitemShut {NoStop}%
\bibitem [{\citenamefont {Wang}\ and\ \citenamefont {Carter}(2002)}]{02Carter}%
  \BibitemOpen
  \bibfield  {author} {\bibinfo {author} {\bibfnamefont {Y.~A.}\ \bibnamefont
  {Wang}}\ and\ \bibinfo {author} {\bibfnamefont {E.~A.}\ \bibnamefont
  {Carter}},\ }\bibfield  {title} {\bibinfo {title} {Orbital-free
  kinetic-energy density functional theory},\ }\href@noop {} {\bibfield
  {journal} {\bibinfo  {journal} {Theoretical Methods in Condensed Phase
  Chemistry}\ ,\ \bibinfo {pages} {117}} (\bibinfo {year} {2002})}\BibitemShut
  {NoStop}%
\bibitem [{\citenamefont {Witt}\ \emph {et~al.}(2018)\citenamefont {Witt},
  \citenamefont {Beatriz}, \citenamefont {Dieterich},\ and\ \citenamefont
  {Carter}}]{18JMR-Witt}%
  \BibitemOpen
  \bibfield  {author} {\bibinfo {author} {\bibfnamefont {W.~C.}\ \bibnamefont
  {Witt}}, \bibinfo {author} {\bibfnamefont {G.}~\bibnamefont {Beatriz}},
  \bibinfo {author} {\bibfnamefont {J.~M.}\ \bibnamefont {Dieterich}},\ and\
  \bibinfo {author} {\bibfnamefont {E.~A.}\ \bibnamefont {Carter}},\ }\bibfield
   {title} {\bibinfo {title} {Orbital-free density functional theory for
  materials research},\ }\href@noop {} {\bibfield  {journal} {\bibinfo
  {journal} {J. Mater. Res.}\ }\textbf {\bibinfo {volume} {33}},\ \bibinfo
  {pages} {777} (\bibinfo {year} {2018})}\BibitemShut {NoStop}%
\bibitem [{\citenamefont {Ho}\ \emph {et~al.}(2008)\citenamefont {Ho},
  \citenamefont {Lign{\`e}res},\ and\ \citenamefont
  {Carter}}]{08CPC-Ho-PROFESS}%
  \BibitemOpen
  \bibfield  {author} {\bibinfo {author} {\bibfnamefont {G.~S.}\ \bibnamefont
  {Ho}}, \bibinfo {author} {\bibfnamefont {V.~L.}\ \bibnamefont
  {Lign{\`e}res}},\ and\ \bibinfo {author} {\bibfnamefont {E.~A.}\ \bibnamefont
  {Carter}},\ }\bibfield  {title} {\bibinfo {title} {Introducing profess: A new
  program for orbital-free density functional theory calculations},\
  }\href@noop {} {\bibfield  {journal} {\bibinfo  {journal} {Comput. Phys.
  Commun.}\ }\textbf {\bibinfo {volume} {179}},\ \bibinfo {pages} {839}
  (\bibinfo {year} {2008})}\BibitemShut {NoStop}%
\bibitem [{\citenamefont {Hung}\ \emph {et~al.}(2010)\citenamefont {Hung},
  \citenamefont {Huang}, \citenamefont {Shin}, \citenamefont {Ho},
  \citenamefont {Lign{\`e}res},\ and\ \citenamefont
  {Carter}}]{10CPC-Hung-PROFESS}%
  \BibitemOpen
  \bibfield  {author} {\bibinfo {author} {\bibfnamefont {L.}~\bibnamefont
  {Hung}}, \bibinfo {author} {\bibfnamefont {C.}~\bibnamefont {Huang}},
  \bibinfo {author} {\bibfnamefont {I.}~\bibnamefont {Shin}}, \bibinfo {author}
  {\bibfnamefont {G.~S.}\ \bibnamefont {Ho}}, \bibinfo {author} {\bibfnamefont
  {V.~L.}\ \bibnamefont {Lign{\`e}res}},\ and\ \bibinfo {author} {\bibfnamefont
  {E.~A.}\ \bibnamefont {Carter}},\ }\bibfield  {title} {\bibinfo {title}
  {Introducing profess 2.0: A parallelized, fully linear scaling program for
  orbital-free density functional theory calculations},\ }\href@noop {}
  {\bibfield  {journal} {\bibinfo  {journal} {Comput. Phys. Commun.}\ }\textbf
  {\bibinfo {volume} {181}},\ \bibinfo {pages} {2208} (\bibinfo {year}
  {2010})}\BibitemShut {NoStop}%
\bibitem [{\citenamefont {Chen}\ \emph {et~al.}(2015)\citenamefont {Chen},
  \citenamefont {Xia}, \citenamefont {Huang}, \citenamefont {Dieterich},
  \citenamefont {Hung}, \citenamefont {Shin},\ and\ \citenamefont
  {Carter}}]{15CPC-Chen-PROFESS}%
  \BibitemOpen
  \bibfield  {author} {\bibinfo {author} {\bibfnamefont {M.}~\bibnamefont
  {Chen}}, \bibinfo {author} {\bibfnamefont {J.}~\bibnamefont {Xia}}, \bibinfo
  {author} {\bibfnamefont {C.}~\bibnamefont {Huang}}, \bibinfo {author}
  {\bibfnamefont {J.~M.}\ \bibnamefont {Dieterich}}, \bibinfo {author}
  {\bibfnamefont {L.}~\bibnamefont {Hung}}, \bibinfo {author} {\bibfnamefont
  {I.}~\bibnamefont {Shin}},\ and\ \bibinfo {author} {\bibfnamefont {E.~A.}\
  \bibnamefont {Carter}},\ }\bibfield  {title} {\bibinfo {title} {Introducing
  profess 3.0: An advanced program for orbital-free density functional theory
  molecular dynamics simulations},\ }\href@noop {} {\bibfield  {journal}
  {\bibinfo  {journal} {Comput. Phys. Commun.}\ }\textbf {\bibinfo {volume}
  {190}},\ \bibinfo {pages} {228} (\bibinfo {year} {2015})}\BibitemShut
  {NoStop}%
\bibitem [{\citenamefont {Mi}\ \emph {et~al.}(2016)\citenamefont {Mi},
  \citenamefont {Shao}, \citenamefont {Su}, \citenamefont {Zhou}, \citenamefont
  {Zhang}, \citenamefont {Li}, \citenamefont {Wang}, \citenamefont {Zhang},
  \citenamefont {Miao}, \citenamefont {Wang} \emph {et~al.}}]{16CPC-Mi-atlas}%
  \BibitemOpen
  \bibfield  {author} {\bibinfo {author} {\bibfnamefont {W.}~\bibnamefont
  {Mi}}, \bibinfo {author} {\bibfnamefont {X.}~\bibnamefont {Shao}}, \bibinfo
  {author} {\bibfnamefont {C.}~\bibnamefont {Su}}, \bibinfo {author}
  {\bibfnamefont {Y.}~\bibnamefont {Zhou}}, \bibinfo {author} {\bibfnamefont
  {S.}~\bibnamefont {Zhang}}, \bibinfo {author} {\bibfnamefont
  {Q.}~\bibnamefont {Li}}, \bibinfo {author} {\bibfnamefont {H.}~\bibnamefont
  {Wang}}, \bibinfo {author} {\bibfnamefont {L.}~\bibnamefont {Zhang}},
  \bibinfo {author} {\bibfnamefont {M.}~\bibnamefont {Miao}}, \bibinfo {author}
  {\bibfnamefont {Y.}~\bibnamefont {Wang}}, \emph {et~al.},\ }\bibfield
  {title} {\bibinfo {title} {Atlas: A real-space finite-difference
  implementation of orbital-free density functional theory},\ }\href@noop {}
  {\bibfield  {journal} {\bibinfo  {journal} {Comput. Phys. Commun.}\ }\textbf
  {\bibinfo {volume} {200}},\ \bibinfo {pages} {87} (\bibinfo {year}
  {2016})}\BibitemShut {NoStop}%
\bibitem [{\citenamefont {Karasiev}\ and\ \citenamefont
  {Trickey}(2012)}]{12CPC-Karasiev}%
  \BibitemOpen
  \bibfield  {author} {\bibinfo {author} {\bibfnamefont {V.~V.}\ \bibnamefont
  {Karasiev}}\ and\ \bibinfo {author} {\bibfnamefont {S.~B.}\ \bibnamefont
  {Trickey}},\ }\bibfield  {title} {\bibinfo {title} {Issues and challenges in
  orbital-free density functional calculations},\ }\href@noop {} {\bibfield
  {journal} {\bibinfo  {journal} {Comput. Phys. Commun.}\ }\textbf {\bibinfo
  {volume} {183}},\ \bibinfo {pages} {2519} (\bibinfo {year}
  {2012})}\BibitemShut {NoStop}%
\bibitem [{\citenamefont {Thomas}(1927)}]{27-Thomas-local}%
  \BibitemOpen
  \bibfield  {author} {\bibinfo {author} {\bibfnamefont {L.~H.}\ \bibnamefont
  {Thomas}},\ }\bibfield  {title} {\bibinfo {title} {The calculation of atomic
  fields},\ }in\ \href@noop {} {\emph {\bibinfo {booktitle} {Mathematical
  proceedings of the Cambridge philosophical society}}},\ Vol.~\bibinfo
  {volume} {23}\ (\bibinfo {organization} {Cambridge University Press},\
  \bibinfo {year} {1927})\ pp.\ \bibinfo {pages} {542--548}\BibitemShut
  {NoStop}%
\bibitem [{\citenamefont {Fermi}(1927)}]{27TANL-Fermi-local}%
  \BibitemOpen
  \bibfield  {author} {\bibinfo {author} {\bibfnamefont {E.}~\bibnamefont
  {Fermi}},\ }\bibfield  {title} {\bibinfo {title} {Statistical method to
  determine some properties of atoms},\ }\href@noop {} {\bibfield  {journal}
  {\bibinfo  {journal} {Rend. Accad. Naz. Lincei}\ }\textbf {\bibinfo {volume}
  {6}},\ \bibinfo {pages} {5} (\bibinfo {year} {1927})}\BibitemShut {NoStop}%
\bibitem [{\citenamefont {Weizs{\"a}cker}(1935)}]{35-vW-semilocal}%
  \BibitemOpen
  \bibfield  {author} {\bibinfo {author} {\bibfnamefont {C.~v.}\ \bibnamefont
  {Weizs{\"a}cker}},\ }\bibfield  {title} {\bibinfo {title} {Zur theorie der
  kernmassen},\ }\href@noop {} {\bibfield  {journal} {\bibinfo  {journal}
  {Zeitschrift f{\"u}r Physik}\ }\textbf {\bibinfo {volume} {96}},\ \bibinfo
  {pages} {431} (\bibinfo {year} {1935})}\BibitemShut {NoStop}%
\bibitem [{\citenamefont {Luo}\ \emph {et~al.}(2018)\citenamefont {Luo},
  \citenamefont {Karasiev},\ and\ \citenamefont {Trickey}}]{18B-Luo-semilocal}%
  \BibitemOpen
  \bibfield  {author} {\bibinfo {author} {\bibfnamefont {K.}~\bibnamefont
  {Luo}}, \bibinfo {author} {\bibfnamefont {V.~V.}\ \bibnamefont {Karasiev}},\
  and\ \bibinfo {author} {\bibfnamefont {S.}~\bibnamefont {Trickey}},\
  }\bibfield  {title} {\bibinfo {title} {A simple generalized gradient
  approximation for the noninteracting kinetic energy density functional},\
  }\href@noop {} {\bibfield  {journal} {\bibinfo  {journal} {Phys. Rev. B}\
  }\textbf {\bibinfo {volume} {98}},\ \bibinfo {pages} {041111} (\bibinfo
  {year} {2018})}\BibitemShut {NoStop}%
\bibitem [{\citenamefont {Constantin}\ \emph {et~al.}(2018)\citenamefont
  {Constantin}, \citenamefont {Fabiano},\ and\ \citenamefont
  {Della~Sala}}]{18JPCL-Constantin-semilocal}%
  \BibitemOpen
  \bibfield  {author} {\bibinfo {author} {\bibfnamefont {L.~A.}\ \bibnamefont
  {Constantin}}, \bibinfo {author} {\bibfnamefont {E.}~\bibnamefont
  {Fabiano}},\ and\ \bibinfo {author} {\bibfnamefont {F.}~\bibnamefont
  {Della~Sala}},\ }\bibfield  {title} {\bibinfo {title} {Semilocal
  pauli--gaussian kinetic functionals for orbital-free density functional
  theory calculations of solids},\ }\href@noop {} {\bibfield  {journal}
  {\bibinfo  {journal} {J. Phys. Chem. Lett.}\ }\textbf {\bibinfo {volume}
  {9}},\ \bibinfo {pages} {4385} (\bibinfo {year} {2018})}\BibitemShut
  {NoStop}%
\bibitem [{\citenamefont {Kang}\ \emph {et~al.}(2020)\citenamefont {Kang},
  \citenamefont {Luo}, \citenamefont {Runge},\ and\ \citenamefont
  {Trickey}}]{20Kang-semilocal}%
  \BibitemOpen
  \bibfield  {author} {\bibinfo {author} {\bibfnamefont {D.}~\bibnamefont
  {Kang}}, \bibinfo {author} {\bibfnamefont {K.}~\bibnamefont {Luo}}, \bibinfo
  {author} {\bibfnamefont {K.}~\bibnamefont {Runge}},\ and\ \bibinfo {author}
  {\bibfnamefont {S.}~\bibnamefont {Trickey}},\ }\bibfield  {title} {\bibinfo
  {title} {Two-temperature warm dense hydrogen as a test of quantum protons
  driven by orbital-free density functional theory electronic forces},\
  }\href@noop {} {\bibfield  {journal} {\bibinfo  {journal} {Matter Radiat. at
  Extremes}\ }\textbf {\bibinfo {volume} {5}},\ \bibinfo {pages} {064403}
  (\bibinfo {year} {2020})}\BibitemShut {NoStop}%
\bibitem [{\citenamefont {Wang}\ and\ \citenamefont
  {Teter}(1992)}]{92B-Wang-nonlocal}%
  \BibitemOpen
  \bibfield  {author} {\bibinfo {author} {\bibfnamefont {L.-W.}\ \bibnamefont
  {Wang}}\ and\ \bibinfo {author} {\bibfnamefont {M.~P.}\ \bibnamefont
  {Teter}},\ }\bibfield  {title} {\bibinfo {title} {Kinetic-energy functional
  of the electron density},\ }\href@noop {} {\bibfield  {journal} {\bibinfo
  {journal} {Phys. Rev. B}\ }\textbf {\bibinfo {volume} {45}},\ \bibinfo
  {pages} {13196} (\bibinfo {year} {1992})}\BibitemShut {NoStop}%
\bibitem [{\citenamefont {Wang}\ \emph {et~al.}(1999)\citenamefont {Wang},
  \citenamefont {Govind},\ and\ \citenamefont {Carter}}]{99B-Wang-nonlocal}%
  \BibitemOpen
  \bibfield  {author} {\bibinfo {author} {\bibfnamefont {Y.~A.}\ \bibnamefont
  {Wang}}, \bibinfo {author} {\bibfnamefont {N.}~\bibnamefont {Govind}},\ and\
  \bibinfo {author} {\bibfnamefont {E.~A.}\ \bibnamefont {Carter}},\ }\bibfield
   {title} {\bibinfo {title} {Orbital-free kinetic-energy density functionals
  with a density-dependent kernel},\ }\href@noop {} {\bibfield  {journal}
  {\bibinfo  {journal} {Phys. Rev. B}\ }\textbf {\bibinfo {volume} {60}},\
  \bibinfo {pages} {16350} (\bibinfo {year} {1999})}\BibitemShut {NoStop}%
\bibitem [{\citenamefont {Huang}\ and\ \citenamefont
  {Carter}(2010)}]{10B-Huang-nonlocal}%
  \BibitemOpen
  \bibfield  {author} {\bibinfo {author} {\bibfnamefont {C.}~\bibnamefont
  {Huang}}\ and\ \bibinfo {author} {\bibfnamefont {E.~A.}\ \bibnamefont
  {Carter}},\ }\bibfield  {title} {\bibinfo {title} {Nonlocal orbital-free
  kinetic energy density functional for semiconductors},\ }\href@noop {}
  {\bibfield  {journal} {\bibinfo  {journal} {Phys. Rev. B}\ }\textbf {\bibinfo
  {volume} {81}},\ \bibinfo {pages} {045206} (\bibinfo {year}
  {2010})}\BibitemShut {NoStop}%
\bibitem [{\citenamefont {Mi}\ \emph {et~al.}(2018)\citenamefont {Mi},
  \citenamefont {Genova},\ and\ \citenamefont {Pavanello}}]{18JCP-Mi-nonlocal}%
  \BibitemOpen
  \bibfield  {author} {\bibinfo {author} {\bibfnamefont {W.}~\bibnamefont
  {Mi}}, \bibinfo {author} {\bibfnamefont {A.}~\bibnamefont {Genova}},\ and\
  \bibinfo {author} {\bibfnamefont {M.}~\bibnamefont {Pavanello}},\ }\bibfield
  {title} {\bibinfo {title} {Nonlocal kinetic energy functionals by functional
  integration},\ }\href@noop {} {\bibfield  {journal} {\bibinfo  {journal} {J.
  Chem. Phys.}\ }\textbf {\bibinfo {volume} {148}},\ \bibinfo {pages} {184107}
  (\bibinfo {year} {2018})}\BibitemShut {NoStop}%
\bibitem [{\citenamefont {Shao}\ \emph {et~al.}(2021)\citenamefont {Shao},
  \citenamefont {Mi},\ and\ \citenamefont {Pavanello}}]{21B-Shao-nonlocal}%
  \BibitemOpen
  \bibfield  {author} {\bibinfo {author} {\bibfnamefont {X.}~\bibnamefont
  {Shao}}, \bibinfo {author} {\bibfnamefont {W.}~\bibnamefont {Mi}},\ and\
  \bibinfo {author} {\bibfnamefont {M.}~\bibnamefont {Pavanello}},\ }\bibfield
  {title} {\bibinfo {title} {Revised huang-carter nonlocal kinetic energy
  functional for semiconductors and their surfaces},\ }\href@noop {} {\bibfield
   {journal} {\bibinfo  {journal} {Phys. Rev. B}\ }\textbf {\bibinfo {volume}
  {104}},\ \bibinfo {pages} {045118} (\bibinfo {year} {2021})}\BibitemShut
  {NoStop}%
\bibitem [{\citenamefont {Huang}\ \emph {et~al.}(2023)\citenamefont {Huang},
  \citenamefont {von Rudorff},\ and\ \citenamefont {von
  Lilienfeld}}]{23Science-Huang-mlfp}%
  \BibitemOpen
  \bibfield  {author} {\bibinfo {author} {\bibfnamefont {B.}~\bibnamefont
  {Huang}}, \bibinfo {author} {\bibfnamefont {G.~F.}\ \bibnamefont {von
  Rudorff}},\ and\ \bibinfo {author} {\bibfnamefont {O.~A.}\ \bibnamefont {von
  Lilienfeld}},\ }\bibfield  {title} {\bibinfo {title} {The central role of
  density functional theory in the ai age},\ }\href@noop {} {\bibfield
  {journal} {\bibinfo  {journal} {Science}\ }\textbf {\bibinfo {volume}
  {381}},\ \bibinfo {pages} {170} (\bibinfo {year} {2023})}\BibitemShut
  {NoStop}%
\bibitem [{\citenamefont {Zhang}\ \emph {et~al.}(2018)\citenamefont {Zhang},
  \citenamefont {Han}, \citenamefont {Wang}, \citenamefont {Car},\ and\
  \citenamefont {Weinan}}]{18prl-zhang-dp}%
  \BibitemOpen
  \bibfield  {author} {\bibinfo {author} {\bibfnamefont {L.}~\bibnamefont
  {Zhang}}, \bibinfo {author} {\bibfnamefont {J.}~\bibnamefont {Han}}, \bibinfo
  {author} {\bibfnamefont {H.}~\bibnamefont {Wang}}, \bibinfo {author}
  {\bibfnamefont {R.}~\bibnamefont {Car}},\ and\ \bibinfo {author}
  {\bibfnamefont {E.}~\bibnamefont {Weinan}},\ }\bibfield  {title} {\bibinfo
  {title} {Deep potential molecular dynamics: a scalable model with the
  accuracy of quantum mechanics},\ }\href@noop {} {\bibfield  {journal}
  {\bibinfo  {journal} {Phys. Rev. Lett.}\ }\textbf {\bibinfo {volume} {120}},\
  \bibinfo {pages} {143001} (\bibinfo {year} {2018})}\BibitemShut {NoStop}%
\bibitem [{\citenamefont {Zhang}\ \emph {et~al.}(2020)\citenamefont {Zhang},
  \citenamefont {Wang}, \citenamefont {Chen}, \citenamefont {Zeng},
  \citenamefont {Zhang}, \citenamefont {Wang},\ and\ \citenamefont
  {Weinan}}]{20CPC-Zhang-dpgen}%
  \BibitemOpen
  \bibfield  {author} {\bibinfo {author} {\bibfnamefont {Y.}~\bibnamefont
  {Zhang}}, \bibinfo {author} {\bibfnamefont {H.}~\bibnamefont {Wang}},
  \bibinfo {author} {\bibfnamefont {W.}~\bibnamefont {Chen}}, \bibinfo {author}
  {\bibfnamefont {J.}~\bibnamefont {Zeng}}, \bibinfo {author} {\bibfnamefont
  {L.}~\bibnamefont {Zhang}}, \bibinfo {author} {\bibfnamefont
  {H.}~\bibnamefont {Wang}},\ and\ \bibinfo {author} {\bibfnamefont
  {E.}~\bibnamefont {Weinan}},\ }\bibfield  {title} {\bibinfo {title} {Dp-gen:
  A concurrent learning platform for the generation of reliable deep learning
  based potential energy models},\ }\href@noop {} {\bibfield  {journal}
  {\bibinfo  {journal} {Comput. Phys. Commun.}\ }\textbf {\bibinfo {volume}
  {253}},\ \bibinfo {pages} {107206} (\bibinfo {year} {2020})}\BibitemShut
  {NoStop}%
\bibitem [{\citenamefont {Chen}\ \emph {et~al.}(2020)\citenamefont {Chen},
  \citenamefont {Zhang}, \citenamefont {Wang},\ and\ \citenamefont
  {E}}]{20JCTC-chen-deepks}%
  \BibitemOpen
  \bibfield  {author} {\bibinfo {author} {\bibfnamefont {Y.}~\bibnamefont
  {Chen}}, \bibinfo {author} {\bibfnamefont {L.}~\bibnamefont {Zhang}},
  \bibinfo {author} {\bibfnamefont {H.}~\bibnamefont {Wang}},\ and\ \bibinfo
  {author} {\bibfnamefont {W.}~\bibnamefont {E}},\ }\bibfield  {title}
  {\bibinfo {title} {Deepks: A comprehensive data-driven approach toward
  chemically accurate density functional theory},\ }\href@noop {} {\bibfield
  {journal} {\bibinfo  {journal} {J. Chem. Theory Comput.}\ }\textbf {\bibinfo
  {volume} {17}},\ \bibinfo {pages} {170} (\bibinfo {year} {2020})}\BibitemShut
  {NoStop}%
\bibitem [{\citenamefont {Kasim}\ and\ \citenamefont
  {Vinko}(2021)}]{21L-Kasim-mlxc}%
  \BibitemOpen
  \bibfield  {author} {\bibinfo {author} {\bibfnamefont {M.~F.}\ \bibnamefont
  {Kasim}}\ and\ \bibinfo {author} {\bibfnamefont {S.~M.}\ \bibnamefont
  {Vinko}},\ }\bibfield  {title} {\bibinfo {title} {Learning the
  exchange-correlation functional from nature with fully differentiable density
  functional theory},\ }\href@noop {} {\bibfield  {journal} {\bibinfo
  {journal} {Phys. Rev. Lett.}\ }\textbf {\bibinfo {volume} {127}},\ \bibinfo
  {pages} {126403} (\bibinfo {year} {2021})}\BibitemShut {NoStop}%
\bibitem [{\citenamefont {Kirkpatrick}\ \emph {et~al.}(2021)\citenamefont
  {Kirkpatrick}, \citenamefont {McMorrow}, \citenamefont {Turban},
  \citenamefont {Gaunt}, \citenamefont {Spencer}, \citenamefont {Matthews},
  \citenamefont {Obika}, \citenamefont {Thiry}, \citenamefont {Fortunato},
  \citenamefont {Pfau} \emph {et~al.}}]{21s-kirkpatrick-dm21}%
  \BibitemOpen
  \bibfield  {author} {\bibinfo {author} {\bibfnamefont {J.}~\bibnamefont
  {Kirkpatrick}}, \bibinfo {author} {\bibfnamefont {B.}~\bibnamefont
  {McMorrow}}, \bibinfo {author} {\bibfnamefont {D.~H.}\ \bibnamefont
  {Turban}}, \bibinfo {author} {\bibfnamefont {A.~L.}\ \bibnamefont {Gaunt}},
  \bibinfo {author} {\bibfnamefont {J.~S.}\ \bibnamefont {Spencer}}, \bibinfo
  {author} {\bibfnamefont {A.~G.}\ \bibnamefont {Matthews}}, \bibinfo {author}
  {\bibfnamefont {A.}~\bibnamefont {Obika}}, \bibinfo {author} {\bibfnamefont
  {L.}~\bibnamefont {Thiry}}, \bibinfo {author} {\bibfnamefont
  {M.}~\bibnamefont {Fortunato}}, \bibinfo {author} {\bibfnamefont
  {D.}~\bibnamefont {Pfau}}, \emph {et~al.},\ }\bibfield  {title} {\bibinfo
  {title} {Pushing the frontiers of density functionals by solving the
  fractional electron problem},\ }\href@noop {} {\bibfield  {journal} {\bibinfo
   {journal} {Science}\ }\textbf {\bibinfo {volume} {374}},\ \bibinfo {pages}
  {1385} (\bibinfo {year} {2021})}\BibitemShut {NoStop}%
\bibitem [{\citenamefont {Nagai}\ \emph {et~al.}(2022)\citenamefont {Nagai},
  \citenamefont {Akashi},\ and\ \citenamefont {Sugino}}]{22PRR-nagai-mlxc}%
  \BibitemOpen
  \bibfield  {author} {\bibinfo {author} {\bibfnamefont {R.}~\bibnamefont
  {Nagai}}, \bibinfo {author} {\bibfnamefont {R.}~\bibnamefont {Akashi}},\ and\
  \bibinfo {author} {\bibfnamefont {O.}~\bibnamefont {Sugino}},\ }\bibfield
  {title} {\bibinfo {title} {Machine-learning-based exchange correlation
  functional with physical asymptotic constraints},\ }\href@noop {} {\bibfield
  {journal} {\bibinfo  {journal} {Phys. Rev. Research}\ }\textbf {\bibinfo
  {volume} {4}},\ \bibinfo {pages} {013106} (\bibinfo {year}
  {2022})}\BibitemShut {NoStop}%
\bibitem [{\citenamefont {Li}\ \emph {et~al.}(2022)\citenamefont {Li},
  \citenamefont {Wang}, \citenamefont {Zou}, \citenamefont {Ye}, \citenamefont
  {Xu}, \citenamefont {Gong}, \citenamefont {Duan},\ and\ \citenamefont
  {Xu}}]{22NCS-Li-deepH}%
  \BibitemOpen
  \bibfield  {author} {\bibinfo {author} {\bibfnamefont {H.}~\bibnamefont
  {Li}}, \bibinfo {author} {\bibfnamefont {Z.}~\bibnamefont {Wang}}, \bibinfo
  {author} {\bibfnamefont {N.}~\bibnamefont {Zou}}, \bibinfo {author}
  {\bibfnamefont {M.}~\bibnamefont {Ye}}, \bibinfo {author} {\bibfnamefont
  {R.}~\bibnamefont {Xu}}, \bibinfo {author} {\bibfnamefont {X.}~\bibnamefont
  {Gong}}, \bibinfo {author} {\bibfnamefont {W.}~\bibnamefont {Duan}},\ and\
  \bibinfo {author} {\bibfnamefont {Y.}~\bibnamefont {Xu}},\ }\bibfield
  {title} {\bibinfo {title} {Deep-learning density functional theory
  hamiltonian for efficient ab initio electronic-structure calculation},\
  }\href@noop {} {\bibfield  {journal} {\bibinfo  {journal} {Nat Comput Sci}\
  }\textbf {\bibinfo {volume} {2}},\ \bibinfo {pages} {367} (\bibinfo {year}
  {2022})}\BibitemShut {NoStop}%
\bibitem [{\citenamefont {Snyder}\ \emph {et~al.}(2012)\citenamefont {Snyder},
  \citenamefont {Rupp}, \citenamefont {Hansen}, \citenamefont {M{\"u}ller},\
  and\ \citenamefont {Burke}}]{12L-Snyder-mlof}%
  \BibitemOpen
  \bibfield  {author} {\bibinfo {author} {\bibfnamefont {J.~C.}\ \bibnamefont
  {Snyder}}, \bibinfo {author} {\bibfnamefont {M.}~\bibnamefont {Rupp}},
  \bibinfo {author} {\bibfnamefont {K.}~\bibnamefont {Hansen}}, \bibinfo
  {author} {\bibfnamefont {K.-R.}\ \bibnamefont {M{\"u}ller}},\ and\ \bibinfo
  {author} {\bibfnamefont {K.}~\bibnamefont {Burke}},\ }\bibfield  {title}
  {\bibinfo {title} {Finding density functionals with machine learning},\
  }\href@noop {} {\bibfield  {journal} {\bibinfo  {journal} {Phys. Rev. Lett.}\
  }\textbf {\bibinfo {volume} {108}},\ \bibinfo {pages} {253002} (\bibinfo
  {year} {2012})}\BibitemShut {NoStop}%
\bibitem [{\citenamefont {Seino}\ \emph {et~al.}(2018)\citenamefont {Seino},
  \citenamefont {Kageyama}, \citenamefont {Fujinami}, \citenamefont {Ikabata},\
  and\ \citenamefont {Nakai}}]{18TJCP-Seino-mlof}%
  \BibitemOpen
  \bibfield  {author} {\bibinfo {author} {\bibfnamefont {J.}~\bibnamefont
  {Seino}}, \bibinfo {author} {\bibfnamefont {R.}~\bibnamefont {Kageyama}},
  \bibinfo {author} {\bibfnamefont {M.}~\bibnamefont {Fujinami}}, \bibinfo
  {author} {\bibfnamefont {Y.}~\bibnamefont {Ikabata}},\ and\ \bibinfo {author}
  {\bibfnamefont {H.}~\bibnamefont {Nakai}},\ }\bibfield  {title} {\bibinfo
  {title} {Semi-local machine-learned kinetic energy density functional with
  third-order gradients of electron density},\ }\href@noop {} {\bibfield
  {journal} {\bibinfo  {journal} {J. Chem. Phys.}\ }\textbf {\bibinfo {volume}
  {148}} (\bibinfo {year} {2018})}\BibitemShut {NoStop}%
\bibitem [{\citenamefont {Hollingsworth}\ \emph {et~al.}(2018)\citenamefont
  {Hollingsworth}, \citenamefont {Li}, \citenamefont {Baker},\ and\
  \citenamefont {Burke}}]{18TJCP-Hollingsworth-mlof}%
  \BibitemOpen
  \bibfield  {author} {\bibinfo {author} {\bibfnamefont {J.}~\bibnamefont
  {Hollingsworth}}, \bibinfo {author} {\bibfnamefont {L.}~\bibnamefont {Li}},
  \bibinfo {author} {\bibfnamefont {T.~E.}\ \bibnamefont {Baker}},\ and\
  \bibinfo {author} {\bibfnamefont {K.}~\bibnamefont {Burke}},\ }\bibfield
  {title} {\bibinfo {title} {Can exact conditions improve machine-learned
  density functionals?},\ }\href@noop {} {\bibfield  {journal} {\bibinfo
  {journal} {J. Chem. Phys.}\ }\textbf {\bibinfo {volume} {148}} (\bibinfo
  {year} {2018})}\BibitemShut {NoStop}%
\bibitem [{\citenamefont {Meyer}\ \emph {et~al.}(2020)\citenamefont {Meyer},
  \citenamefont {Weichselbaum},\ and\ \citenamefont
  {Hauser}}]{20JCTC-Meyer-mlof}%
  \BibitemOpen
  \bibfield  {author} {\bibinfo {author} {\bibfnamefont {R.}~\bibnamefont
  {Meyer}}, \bibinfo {author} {\bibfnamefont {M.}~\bibnamefont
  {Weichselbaum}},\ and\ \bibinfo {author} {\bibfnamefont {A.~W.}\ \bibnamefont
  {Hauser}},\ }\bibfield  {title} {\bibinfo {title} {Machine learning
  approaches toward orbital-free density functional theory: Simultaneous
  training on the kinetic energy density functional and its functional
  derivative},\ }\href@noop {} {\bibfield  {journal} {\bibinfo  {journal} {J.
  Chem. Theory Comput.}\ }\textbf {\bibinfo {volume} {16}},\ \bibinfo {pages}
  {5685} (\bibinfo {year} {2020})}\BibitemShut {NoStop}%
\bibitem [{\citenamefont {Imoto}\ \emph {et~al.}(2021)\citenamefont {Imoto},
  \citenamefont {Imada},\ and\ \citenamefont {Oshiyama}}]{21PRR-Imoto-mlof}%
  \BibitemOpen
  \bibfield  {author} {\bibinfo {author} {\bibfnamefont {F.}~\bibnamefont
  {Imoto}}, \bibinfo {author} {\bibfnamefont {M.}~\bibnamefont {Imada}},\ and\
  \bibinfo {author} {\bibfnamefont {A.}~\bibnamefont {Oshiyama}},\ }\bibfield
  {title} {\bibinfo {title} {Order-n orbital-free density-functional
  calculations with machine learning of functional derivatives for
  semiconductors and metals},\ }\href@noop {} {\bibfield  {journal} {\bibinfo
  {journal} {Phys. Rev. Research}\ }\textbf {\bibinfo {volume} {3}},\ \bibinfo
  {pages} {033198} (\bibinfo {year} {2021})}\BibitemShut {NoStop}%
\bibitem [{\citenamefont {Ryczko}\ \emph {et~al.}(2022)\citenamefont {Ryczko},
  \citenamefont {Wetzel}, \citenamefont {Melko},\ and\ \citenamefont
  {Tamblyn}}]{22JCTC-Ryczko-mlof}%
  \BibitemOpen
  \bibfield  {author} {\bibinfo {author} {\bibfnamefont {K.}~\bibnamefont
  {Ryczko}}, \bibinfo {author} {\bibfnamefont {S.~J.}\ \bibnamefont {Wetzel}},
  \bibinfo {author} {\bibfnamefont {R.~G.}\ \bibnamefont {Melko}},\ and\
  \bibinfo {author} {\bibfnamefont {I.}~\bibnamefont {Tamblyn}},\ }\bibfield
  {title} {\bibinfo {title} {Toward orbital-free density functional theory with
  small data sets and deep learning},\ }\href@noop {} {\bibfield  {journal}
  {\bibinfo  {journal} {J. Chem. Theory Comput.}\ }\textbf {\bibinfo {volume}
  {18}},\ \bibinfo {pages} {1122} (\bibinfo {year} {2022})}\BibitemShut
  {NoStop}%
\bibitem [{\citenamefont {Mazo-Sevillano}\ and\ \citenamefont
  {Hermann}(2023)}]{23JCTC-Pablo}%
  \BibitemOpen
  \bibfield  {author} {\bibinfo {author} {\bibfnamefont {P.~d.}\ \bibnamefont
  {Mazo-Sevillano}}\ and\ \bibinfo {author} {\bibfnamefont {J.}~\bibnamefont
  {Hermann}},\ }\bibfield  {title} {\bibinfo {title} {Variational principle to
  regularize machine-learned density functionals: The non-interacting
  kinetic-energy functional},\ }\href@noop {} {\bibfield  {journal} {\bibinfo
  {journal} {J. Chem. Theory Comput.}\ }\textbf {\bibinfo {volume} {159}}
  (\bibinfo {year} {2023})}\BibitemShut {NoStop}%
\bibitem [{\citenamefont {Levy}\ and\ \citenamefont
  {Ou-Yang}(1988)}]{88PRA-Levy-pauli}%
  \BibitemOpen
  \bibfield  {author} {\bibinfo {author} {\bibfnamefont {M.}~\bibnamefont
  {Levy}}\ and\ \bibinfo {author} {\bibfnamefont {H.}~\bibnamefont {Ou-Yang}},\
  }\bibfield  {title} {\bibinfo {title} {Exact properties of the pauli
  potential for the square root of the electron density and the kinetic energy
  functional},\ }\href@noop {} {\bibfield  {journal} {\bibinfo  {journal}
  {Phys. Rev. A}\ }\textbf {\bibinfo {volume} {38}},\ \bibinfo {pages} {625}
  (\bibinfo {year} {1988})}\BibitemShut {NoStop}%
\bibitem [{SM()}]{SM}%
  \BibitemOpen
  \bibinfo {note} {See Supplemental Material at [URL] for the detailed
  parameters used in OFDFT and KSDFT calculations, the formula of Pauli
  potential of the MPN KEDF, the derivation of the FEG limit of Pauli potential
  and the scaling invariance of descriptors, the details of alloy testing set,
  and the test result on random Mg-Al alloys.}\BibitemShut {Stop}%
\bibitem [{\citenamefont {Carling}\ and\ \citenamefont
  {Carter}(2003)}]{03MSMSE-Carling-mgal}%
  \BibitemOpen
  \bibfield  {author} {\bibinfo {author} {\bibfnamefont {K.~M.}\ \bibnamefont
  {Carling}}\ and\ \bibinfo {author} {\bibfnamefont {E.~A.}\ \bibnamefont
  {Carter}},\ }\bibfield  {title} {\bibinfo {title} {Orbital-free density
  functional theory calculations of the properties of al, mg and al--mg
  crystalline phases},\ }\href@noop {} {\bibfield  {journal} {\bibinfo
  {journal} {Model. Simul. Mater. Sci. Eng.}\ }\textbf {\bibinfo {volume}
  {11}},\ \bibinfo {pages} {339} (\bibinfo {year} {2003})}\BibitemShut
  {NoStop}%
\bibitem [{\citenamefont {Jain}\ \emph {et~al.}(2013)\citenamefont {Jain},
  \citenamefont {Ong}, \citenamefont {Hautier}, \citenamefont {Chen},
  \citenamefont {Richards}, \citenamefont {Dacek}, \citenamefont {Cholia},
  \citenamefont {Gunter}, \citenamefont {Skinner}, \citenamefont {Ceder} \emph
  {et~al.}}]{13APL-Jain-MP}%
  \BibitemOpen
  \bibfield  {author} {\bibinfo {author} {\bibfnamefont {A.}~\bibnamefont
  {Jain}}, \bibinfo {author} {\bibfnamefont {S.~P.}\ \bibnamefont {Ong}},
  \bibinfo {author} {\bibfnamefont {G.}~\bibnamefont {Hautier}}, \bibinfo
  {author} {\bibfnamefont {W.}~\bibnamefont {Chen}}, \bibinfo {author}
  {\bibfnamefont {W.~D.}\ \bibnamefont {Richards}}, \bibinfo {author}
  {\bibfnamefont {S.}~\bibnamefont {Dacek}}, \bibinfo {author} {\bibfnamefont
  {S.}~\bibnamefont {Cholia}}, \bibinfo {author} {\bibfnamefont
  {D.}~\bibnamefont {Gunter}}, \bibinfo {author} {\bibfnamefont
  {D.}~\bibnamefont {Skinner}}, \bibinfo {author} {\bibfnamefont
  {G.}~\bibnamefont {Ceder}}, \emph {et~al.},\ }\bibfield  {title} {\bibinfo
  {title} {Commentary: The materials project: A materials genome approach to
  accelerating materials innovation},\ }\href@noop {} {\bibfield  {journal}
  {\bibinfo  {journal} {APL materials}\ }\textbf {\bibinfo {volume} {1}}
  (\bibinfo {year} {2013})}\BibitemShut {NoStop}%
\bibitem [{\citenamefont {Li}\ \emph {et~al.}(2016)\citenamefont {Li},
  \citenamefont {Liu}, \citenamefont {Chen}, \citenamefont {Lin}, \citenamefont
  {Ren}, \citenamefont {Lin}, \citenamefont {Yang},\ and\ \citenamefont
  {He}}]{16Li-CMS-ABACUS}%
  \BibitemOpen
  \bibfield  {author} {\bibinfo {author} {\bibfnamefont {P.}~\bibnamefont
  {Li}}, \bibinfo {author} {\bibfnamefont {X.}~\bibnamefont {Liu}}, \bibinfo
  {author} {\bibfnamefont {M.}~\bibnamefont {Chen}}, \bibinfo {author}
  {\bibfnamefont {P.}~\bibnamefont {Lin}}, \bibinfo {author} {\bibfnamefont
  {X.}~\bibnamefont {Ren}}, \bibinfo {author} {\bibfnamefont {L.}~\bibnamefont
  {Lin}}, \bibinfo {author} {\bibfnamefont {C.}~\bibnamefont {Yang}},\ and\
  \bibinfo {author} {\bibfnamefont {L.}~\bibnamefont {He}},\ }\bibfield
  {title} {\bibinfo {title} {Large-scale ab initio simulations based on
  systematically improvable atomic basis},\ }\href@noop {} {\bibfield
  {journal} {\bibinfo  {journal} {Comp. Mater. Sci.}\ }\textbf {\bibinfo
  {volume} {112}},\ \bibinfo {pages} {503} (\bibinfo {year}
  {2016})}\BibitemShut {NoStop}%
\bibitem [{\citenamefont {Paszke}\ \emph {et~al.}(2019)\citenamefont {Paszke},
  \citenamefont {Gross}, \citenamefont {Massa}, \citenamefont {Lerer},
  \citenamefont {Bradbury}, \citenamefont {Chanan}, \citenamefont {Killeen},
  \citenamefont {Lin}, \citenamefont {Gimelshein}, \citenamefont {Antiga} \emph
  {et~al.}}]{19ANIPS-paszke-pytorch}%
  \BibitemOpen
  \bibfield  {author} {\bibinfo {author} {\bibfnamefont {A.}~\bibnamefont
  {Paszke}}, \bibinfo {author} {\bibfnamefont {S.}~\bibnamefont {Gross}},
  \bibinfo {author} {\bibfnamefont {F.}~\bibnamefont {Massa}}, \bibinfo
  {author} {\bibfnamefont {A.}~\bibnamefont {Lerer}}, \bibinfo {author}
  {\bibfnamefont {J.}~\bibnamefont {Bradbury}}, \bibinfo {author}
  {\bibfnamefont {G.}~\bibnamefont {Chanan}}, \bibinfo {author} {\bibfnamefont
  {T.}~\bibnamefont {Killeen}}, \bibinfo {author} {\bibfnamefont
  {Z.}~\bibnamefont {Lin}}, \bibinfo {author} {\bibfnamefont {N.}~\bibnamefont
  {Gimelshein}}, \bibinfo {author} {\bibfnamefont {L.}~\bibnamefont {Antiga}},
  \emph {et~al.},\ }\bibfield  {title} {\bibinfo {title} {Pytorch: An
  imperative style, high-performance deep learning library},\ }\href@noop {}
  {\bibfield  {journal} {\bibinfo  {journal} {Advances in neural information
  processing systems}\ }\textbf {\bibinfo {volume} {32}} (\bibinfo {year}
  {2019})}\BibitemShut {NoStop}%
\bibitem [{\citenamefont {Monkhorst}\ and\ \citenamefont
  {Pack}(1976)}]{76B-Monkhorst}%
  \BibitemOpen
  \bibfield  {author} {\bibinfo {author} {\bibfnamefont {H.~J.}\ \bibnamefont
  {Monkhorst}}\ and\ \bibinfo {author} {\bibfnamefont {J.~D.}\ \bibnamefont
  {Pack}},\ }\bibfield  {title} {\bibinfo {title} {Special points for
  brillouin-zone integrations},\ }\href@noop {} {\bibfield  {journal} {\bibinfo
   {journal} {Phys. Rev. B}\ }\textbf {\bibinfo {volume} {13}},\ \bibinfo
  {pages} {5188} (\bibinfo {year} {1976})}\BibitemShut {NoStop}%
\bibitem [{\citenamefont {Perdew}\ \emph {et~al.}(1996)\citenamefont {Perdew},
  \citenamefont {Burke},\ and\ \citenamefont {Ernzerhof}}]{96PRL-Perdew-PBE}%
  \BibitemOpen
  \bibfield  {author} {\bibinfo {author} {\bibfnamefont {J.~P.}\ \bibnamefont
  {Perdew}}, \bibinfo {author} {\bibfnamefont {K.}~\bibnamefont {Burke}},\ and\
  \bibinfo {author} {\bibfnamefont {M.}~\bibnamefont {Ernzerhof}},\ }\bibfield
  {title} {\bibinfo {title} {Generalized gradient approximation made simple},\
  }\href@noop {} {\bibfield  {journal} {\bibinfo  {journal} {Phys. Rev. Lett.}\
  }\textbf {\bibinfo {volume} {77}},\ \bibinfo {pages} {3865} (\bibinfo {year}
  {1996})}\BibitemShut {NoStop}%
\bibitem [{\citenamefont {Huang}\ and\ \citenamefont
  {Carter}(2008)}]{08PCCP-Huang-BLPS}%
  \BibitemOpen
  \bibfield  {author} {\bibinfo {author} {\bibfnamefont {C.}~\bibnamefont
  {Huang}}\ and\ \bibinfo {author} {\bibfnamefont {E.~A.}\ \bibnamefont
  {Carter}},\ }\bibfield  {title} {\bibinfo {title} {Transferable local
  pseudopotentials for magnesium, aluminum and silicon},\ }\href@noop {}
  {\bibfield  {journal} {\bibinfo  {journal} {Phys. Chem. Chem. Phys.}\
  }\textbf {\bibinfo {volume} {10}},\ \bibinfo {pages} {7109} (\bibinfo {year}
  {2008})}\BibitemShut {NoStop}%
\bibitem [{\citenamefont {Murnaghan}(1944)}]{44-Murnaghan-bulkmodulus}%
  \BibitemOpen
  \bibfield  {author} {\bibinfo {author} {\bibfnamefont {F.}~\bibnamefont
  {Murnaghan}},\ }\bibfield  {title} {\bibinfo {title} {The compressibility of
  media under extreme pressures},\ }\href@noop {} {\bibfield  {journal}
  {\bibinfo  {journal} {Proc. Natl. Acad. Sci.}\ }\textbf {\bibinfo {volume}
  {30}},\ \bibinfo {pages} {244} (\bibinfo {year} {1944})}\BibitemShut
  {NoStop}%
\bibitem [{\citenamefont {Berk}(1983)}]{83pra-berk-semilocal}%
  \BibitemOpen
  \bibfield  {author} {\bibinfo {author} {\bibfnamefont {A.}~\bibnamefont
  {Berk}},\ }\bibfield  {title} {\bibinfo {title} {Lower-bound energy
  functionals and their application to diatomic systems},\ }\href@noop {}
  {\bibfield  {journal} {\bibinfo  {journal} {Phys. Rev. A}\ }\textbf {\bibinfo
  {volume} {28}},\ \bibinfo {pages} {1908} (\bibinfo {year}
  {1983})}\BibitemShut {NoStop}%
\bibitem [{\citenamefont {Sun}\ \emph {et~al.}(2023)\citenamefont {Sun},
  \citenamefont {Li},\ and\ \citenamefont {Chen}}]{23B-Sun-TKK}%
  \BibitemOpen
  \bibfield  {author} {\bibinfo {author} {\bibfnamefont {L.}~\bibnamefont
  {Sun}}, \bibinfo {author} {\bibfnamefont {Y.}~\bibnamefont {Li}},\ and\
  \bibinfo {author} {\bibfnamefont {M.}~\bibnamefont {Chen}},\ }\bibfield
  {title} {\bibinfo {title} {Truncated nonlocal kinetic energy density
  functionals for simple metals and silicon},\ }\href@noop {} {\bibfield
  {journal} {\bibinfo  {journal} {Phys. Rev. B}\ }\textbf {\bibinfo {volume}
  {108}},\ \bibinfo {pages} {075158} (\bibinfo {year} {2023})}\BibitemShut
  {NoStop}%
\end{thebibliography}%
\end{document}



\title{Supplemental Material: \\ Machine learning based nonlocal kinetic energy density functional for simple metals and alloys}

\author{Liang Sun}
\affiliation{HEDPS, CAPT, School of Physics and College of Engineering, Peking University, Beijing 100871, P. R. China}
\author{Mohan Chen}
\email{mohanchen@pku.edu.cn}
\affiliation{HEDPS, CAPT, School of Physics and College of Engineering, Peking University, Beijing 100871, P. R. China}
\affiliation{AI for Science Institute, Beijing 100080, P. R. China}
\date{\today}
\pacs{71.15.Mb, 07.05.Mh, 71.20.Gj}

\maketitle

\section{Parameters used in OFDFT and KSDFT calculations}

The detailed parameters of KSDFT and OFDFT calculations are shown in Table~\ref{tab:parameters}, including the energy cutoffs and the $k$-point samplings of KSDFT calculations, as well as the energy cutoffs used in OFDFT.

\begin{table}[ht]
    \centering
    \caption{Energy cutoff ($E_{\tx{cut}}$ in eV) and $k$-point mesh of KSDFT, and energy cutoff ($E_{\tx{cut}}$ in eV) of OFDFT. For alloys, the Energy cutoff of both KSDFT and OFDFT were set to 800 eV, and the $k$-point mesh of KSDFT were set to ensure the smallest spacing between k points is not larger than 0.05 $\rm{Bohr}^{-1}$}
    \begin{tabularx}{0.45\linewidth}{
    >{\raggedright\arraybackslash\hsize=1.5\hsize\linewidth=\hsize}X
    >{\centering\arraybackslash\hsize=.75\hsize\linewidth=\hsize}X
    >{\centering\arraybackslash\hsize=.75\hsize\linewidth=\hsize}X
    >{\centering\arraybackslash\hsize=1\hsize\linewidth=\hsize}X}
    \hline\hline
                                           &OFDFT   &KSDFT  &KSDFT\\
        System                             &$E_{\tx{cut}}$&$E_{\tx{cut}}$&$k$-point mesh\\\hline
        fcc, bcc, sc Al                    &$800$&$800$&$20 \times 20 \times 20$\\
        hcp Al, hcp Mg                     &$800$&$800$&$12 \times 12 \times 12$\\
        bcc, fcc, sc, CD Li                &$800$&$800$&$20 \times 20 \times 20$\\
        fcc, bcc, sc Mg                    &$800$&$800$&$20 \times 20 \times 20$\\
        \hline\hline
    \end{tabularx}
    \label{tab:parameters}
\end{table}

\section{Pauli potential of MPN KEDF}

The Puali energy of MPN KEDF is defined as
\begin{equation}
    T^{\rm{MPN}}_\theta = \int{\tau_{\rm{TF}}(r)} F_{\rm{\theta}}^{\rm{NN}} \left(\Tilde{p}(r), \Tilde{p}_{\rm{nl}}(r), \Tilde{\xi}(r), \Tilde{\xi}_{\rm{nl}}(r)\right) {\rm{d}}^3{r},
\end{equation}
where $\tau_{\rm{TF}} = \frac{3}{10}(3\pi^2)^{2/3} \rho^{5/3}$ denotes the Thomas-Fermi (TF) kinetic energy density~\cite{27-Thomas-local, 27TANL-Fermi-local}, and $\Tilde{p}(r) = \tanh{\Big(\chi_p p(r)\Big)}, \Tilde{p}_{\rm{nl}}(r) = \int{w(r-r')\Tilde{p}(r'){\rm{d}}^3 r'}, \Tilde{\xi}(r) = \tanh{\Big(\xi(r)\Big)}, \Tilde{\xi}_{\rm{nl}}(r) = \int{w(r-r')\Tilde{\xi}(r'){\rm{d}}^3 r'}$ are four descriptors introduced in Sec. II D, with $p(r) = |\nabla \rho(r)|^2 / \Big(2(3\pi^2)^{1/3} \rho^{4/3}(r)\Big)^2, \xi(r)=\frac{\int{w(r-r')\rho^{1/3}(r'){\rm{d}}^3 r'}}{\rho^{1/3}(r)}$, and $\chi_p$ is a hyper-parameter and is set to 0.2 in this work.

As a result, the Pauli potential of MPN KEDF is given by
\begin{equation}
    \begin{aligned}
    V_{\theta}^{\rm{MPN}}(r)
    =&
    \frac{\tau_{\rm{TF}}}{\rho}\left(
    \frac{5}{3}F_{\rm{\theta}}^{\rm{NN}}
    -\frac{8}{3}\chi_{p}p(1-\tilde{p}^2)\frac{\partial F_{\rm{\theta}}^{\rm{NN}}}{\partial \tilde{p}}
    -\frac{1}{3}\xi(1-\tilde{\xi}^2)\frac{\partial F_{\rm{\theta}}^{\rm{NN}}}{\partial\tilde{\xi}} \right)\\
    &
    -\frac{3}{20}\chi_{p}\nabla\cdot\left((1-\tilde{p}^2)\frac{\nabla\rho}{\rho}\frac{\partial F_{\rm{\theta}}^{\rm{NN}}}{\partial\tilde{p}}\right)\\
    &
    -\frac{8}{3}\chi_{p}(1-\tilde{p}^2)\frac{p}{ \rho}\int{w(r'-r)\tau_{\rm{TF}}(r')\frac{\partial F_{\rm{\theta}}^{\rm{NN}}}{\partial \tilde{p}_{\rm{nl}}}{\rm{d}}^3{r}'}\\
    &
    -\chi_{p}\nabla\cdot\left((1-\tilde{p}^2)\frac{\nabla\rho}{2(3\pi^2)^{2/3} \rho^{8/3}}\int{w(r'-r)\tau_{\rm{TF}}(r')\frac{\partial F_{\rm{\theta}}^{\rm{NN}}}{\partial \tilde{p}_{\rm{nl}}}}{\rm{d}}^3{r'}\right)\\
    &
    +\frac{1}{3}\frac{1}{\rho^{2/3}}\int{w(r'-r)\frac{\tau_{\rm{TF}}(r')}{\rho^{1/3}(r')}(1-\tilde{\xi}^2(r'))\frac{\partial F_{\rm{\theta}}^{\rm{NN}}}{\partial\tilde{\xi}} {\rm{d}}^3{r'}}\\
    &
    -\frac{1}{3}\frac{\xi}{\rho}(1-\tilde{\xi}^2)\int{w(r'-r)\tau_{\rm{TF}}(r')\frac{\partial F_{\rm{\theta}}^{\rm{NN}}}{\partial\tilde{\xi}_{\rm{nl}}}{\rm{d}}^3{r'}}\\
    &
    +\frac{1}{3}\frac{1}{\rho^{2/3}}\int{w(r''-r)\frac{1-\tilde{\xi}^2(r'')}{\rho^{1/3}(r'')}\int{w(r'-r'')\tau_{\rm{TF}}(r')\frac{\partial F_{\rm{\theta}}^{\rm{NN}}}{\partial\tilde{\xi}_{\rm{nl}}}{\rm{d}}^3{r'}}{\rm{d}}^3{r''}},
    \end{aligned}
    \label{eq.mpn_p}
\end{equation}
where ${\partial F_{\rm{\theta}}^{\rm{NN}}}/{\partial \tilde{p}}, {\partial F_{\rm{\theta}}^{\rm{NN}}}/{\partial \tilde{p}_{\rm{nl}}}, {\partial F_{\rm{\theta}}^{\rm{NN}}}/{\partial \tilde{\xi}}, {\partial F_{\rm{\theta}}^{\rm{NN}}}/{\partial \tilde{\xi}_{\rm{nl}}}$ can be obtained through the back propagation of NN.
The detailed derivation will be shown below.

First of all, we have
\begin{equation}
    \begin{aligned}
    V_{\theta}^{\rm{MPN}}(r) =& \frac{\delta T_{\theta}^{\rm{MPN}}}{\delta \rho(r)} \\
    =&
    \frac{\delta}{\delta\rho(r)}\int{\tau_{\rm{TF}}(r')F_{\rm{\theta}}^{\rm{NN}} \left(\Tilde{p}(r'), \Tilde{p}_{\rm{nl}}(r'), \Tilde{\xi}(r'), \Tilde{\xi}_{\rm{nl}}(r')\right){\rm{d}}^3{r'}}\\
    =&
    \frac{5}{3}\frac{\tau_{\rm{TF}}F_{\rm{\theta}}^{\rm{NN}}}{\rho} 
    + \int{\tau_{\rm{TF}}\frac{\partial F_{\rm{\theta}}^{\rm{NN}}}{\partial \tilde{p}}\frac{\delta \tilde{p}}{\delta \rho}{\rm{d}}^3{r'}}
    + \int{\tau_{\rm{TF}}\frac{\partial F_{\rm{\theta}}^{\rm{NN}}}{\partial \tilde{p}_{\rm{nl}}}\frac{\delta \tilde{p}_{\rm{nl}}}{\delta \rho}{\rm{d}}^3{r'}}
    + \int{\tau_{\rm{TF}}\frac{\partial F_{\rm{\theta}}^{\rm{NN}}}{\partial \tilde{\xi}}\frac{\delta \tilde{\xi}}{\delta \rho}{\rm{d}}^3{r'}}
    + \int{\tau_{\rm{TF}}\frac{\partial F_{\rm{\theta}}^{\rm{NN}}}{\partial \tilde{\xi}_{\rm{nl}}}\frac{\delta \tilde{\xi}_{\rm{nl}}}{\delta \rho}{\rm{d}}^3{r'}}.
    \end{aligned}
\end{equation}
The first term is explicitly stated, and we shall proceed to derive the remaining four terms in a step-by-step manner.

The second term is introduced by $\tilde{p}$, and it gives
\begin{equation}
    \int{\tau_{\rm{TF}}\frac{\partial F_{\rm{\theta}}^{\rm{NN}}}{\partial \tilde{p}}\frac{\delta \tilde{p}}{\delta \rho}{\rm{d}}^3{r'}}
    = \int{\tau_{\rm{TF}}\frac{\partial F_{\rm{\theta}}^{\rm{NN}}}{\partial \tilde{p}}\frac{\partial \tilde{p}(r')}{\partial p(r')}\frac{\delta p(r')}{\delta \rho(r)}{\rm{d}}^3{r'}}
    = \int{\tau_{\rm{TF}}\frac{\partial F_{\rm{\theta}}^{\rm{NN}}}{\partial \tilde{p}}\chi_{p}(1-\tilde{p}^2(r'))\frac{\delta p(r')}{\delta \rho(r)}{\rm{d}}^3{r'}},
\end{equation}
we note that ${{\rm{d}} \tanh(\chi x)}/{{\rm{d}} x} = \chi(1 - \tanh^2(\chi{x}))$.
Taking into account the expression
\begin{equation}
    \begin{aligned}
    \frac{\delta p(r')}{\delta \rho(r)} 
    &
    = \frac{\partial p(r')}{\partial \rho(r')}\frac{\delta \rho(r')}{\delta \rho(r)} + \frac{\partial p(r')}{\partial |\nabla_{r'}\rho(r')|^2}\frac{\delta |\nabla_{r'}\rho(r')|^2}{\delta \rho(r)}\\
    &
    = -\frac{8}{3}\frac{p(r')}{ \rho(r')}\delta(r-r')
    + \frac{1}{4(3\pi^2)^{2/3} \rho^{8/3}}2\nabla_{r'}\rho(r')\cdot\nabla_{r'}\left(\delta(r-r')\right),
    \end{aligned}
\end{equation}
and utilizing the first Green’s identities $\int{\bm{v}(r)\cdot \nabla f(r){\rm{d}}^3{r}} = - \int{f(r)\nabla \cdot \bm{v}(r) {\rm{d}}^3{r}}$, we obtain
\begin{equation}
    \int{\tau_{\rm{TF}}\frac{\partial F_{\rm{\theta}}^{\rm{NN}}}{\partial \tilde{p}}\frac{\delta \tilde{p}}{\delta \rho}{\rm{d}}^3{r'}}
    = -\frac{8}{3}\chi_{p} \frac{\tau_{\rm{TF}}}{\rho} p(1-\tilde{p}^2)\frac{\partial F_{\rm{\theta}}^{\rm{NN}}}{\partial \tilde{p}}
    -\frac{3}{20}\chi_{p}\nabla\cdot\left((1-\tilde{p}^2)\frac{\nabla\rho}{\rho}\frac{\partial F_{\rm{\theta}}^{\rm{NN}}}{\partial\tilde{p}}\right).
\end{equation}

In order to derive the third term introduced by $\tilde{p}_{\rm{nl}}$, we first calculate ${\delta \tilde{p}_{\rm{nl}}(r')}/{\delta \rho(r)}$ as follows:
\begin{equation}
    \begin{aligned}
    \frac{\delta \tilde{p}_{\rm{nl}}(r')}{\delta \rho(r)}
    &
    = \int{w(r'-r'')\frac{\partial \tilde{p}(r'')}{\partial p(r'')}\frac{\delta p(r'')}{\delta \rho(r)}{\rm{d}}^3{r''}}\\
    &
    = - \frac{8}{3}\chi_{p}(1-\tilde{p}^2)w(r'-r)\frac{p(r)}{ \rho(r)}
    - \chi_{p}\nabla\cdot\left((1-\tilde{p}^2(r))\frac{w(r'-r)}{2(3\pi^2)^{2/3} \rho^{8/3}(r)}\nabla\rho(r)\right).
    \end{aligned}
\end{equation}
Consequently, the third term can be expressed as:
\begin{equation}
    \begin{aligned}
    \int{\tau_{\rm{TF}}\frac{\partial F_{\rm{\theta}}^{\rm{NN}}}{\partial \tilde{p}_{\rm{nl}}}\frac{\delta \tilde{p}_{\rm{nl}}}{\delta \rho}{\rm{d}}^3{r'}} 
    =&
    -\frac{8}{3}\chi_{p}(1-\tilde{p}^2)\frac{p(r)}{ \rho(r)}\int{w(r'-r)\tau_{\rm{TF}}(r')\frac{\partial F_{\rm{\theta}}^{\rm{NN}}}{\partial \tilde{p}_{\rm{nl}}}{\rm{d}}^3{r'}}\\
    &
    -\chi_{p}\nabla\cdot\left((1-\tilde{p}^2(r))\frac{\nabla\rho(r)}{2(3\pi^2)^{2/3} \rho^{8/3}(r)}\int{w(r'-r)\tau_{\rm{TF}}(r')\frac{\partial F_{\rm{\theta}}^{\rm{NN}}}{\partial \tilde{p}_{\rm{nl}}}}{\rm{d}}^3{r'}\right).
    \end{aligned}
\end{equation}

In order to derive the fourth term corresponding to $\tilde{\xi} = \tanh{(\xi)}$, we first define $\gamma = \rho^{1/3}$, and $\gamma_{\rm{nl}} = \int{w(r-r')\gamma(r'){\rm{d}}^3 r'}$, so that $\xi = {\gamma_{\rm{nl}}}/{\gamma}$. Then, we have
\begin{equation}
    \begin{aligned}
    \frac{\delta \tilde{\xi}(r')}{\delta \rho(r)}
    &
    = \frac{\partial \tilde{\xi}(r')}{\partial \xi(r')} \frac{\delta \xi(r')}{\delta \rho(r)}\\
    &
    = (1 - \tilde{\xi}^2(r')) 
    \left(\frac{\partial \xi(r')}{\partial \gamma(r')}\frac{\delta \gamma(r')}{\delta \rho(r)} 
    +\frac{\partial \xi(r')}{\partial \gamma_{\rm{nl}}(r')}\frac{\delta \gamma_{\rm{nl}}(r')}{\delta \rho(r)}\right)\\
    &
    = (1 - \tilde{\xi}^2(r')) 
    \left(- \frac{\xi(r')}{\gamma(r')}\frac{1}{3}\frac{\gamma(r')}{\rho(r')}\delta(r-r') 
    +\frac{1}{\gamma(r')}\frac{1}{3}w(r'-r)\frac{\gamma(r)}{\rho(r)}\right).
    \end{aligned}
\end{equation}
Consequently, the fourth term can be expressed as follows:
\begin{equation}
    \begin{aligned}
    \int{\tau_{\rm{TF}}\frac{\partial F_{\rm{\theta}}^{\rm{NN}}}{\partial \tilde{\xi}}\frac{\delta \tilde{\xi}}{\delta \rho}{\rm{d}}^3{r'}}
    =&
    -\frac{1}{3}\frac{\tau_{\rm{TF}}}{\rho}\xi(1-\tilde{\xi}^2)\frac{\partial F_{\rm{\theta}}^{\rm{NN}}}{\partial\tilde{\xi}}\\
    &
    +\frac{1}{3}\frac{1}{\rho^{2/3}}\int{w(r'-r)\frac{\tau_{\rm{TF}}(r')}{\rho^{1/3}(r')}(1-\tilde{\xi}^2(r'))\frac{\partial F_{\rm{\theta}}^{\rm{NN}}}{\partial\tilde{\xi}} {\rm{d}}^3{r'}}.
    \end{aligned}
\end{equation}

The last term is introduced by $\Tilde{\xi}_{\rm{nl}}(r) = \int{w(r-r')\Tilde{\xi}(r'){\rm{d}}^3 r'}$, and we have
\begin{equation}
    \begin{aligned}
    \int{\tau_{\rm{TF}}\frac{\partial F_{\rm{\theta}}^{\rm{NN}}}{\partial \tilde{\xi}}\frac{\delta \tilde{\xi}_{\rm{nl}}}{\delta \rho}{\rm{d}}^3{r'}}
    =&
    \int{\tau_{\rm{TF}}\frac{\partial F_{\rm{\theta}}^{\rm{NN}}}{\partial \tilde{\xi}_{\rm{nl}}}\int{w(r'-r'')\frac{\partial \tilde{\xi}(r'')}{\partial \xi(r'')}\left(\frac{\partial \xi(r'')}{\partial \gamma(r'')}\frac{\delta \gamma(r'')}{\delta \rho(r)} + \frac{\partial \xi(r'')}{\partial \gamma_{\rm{nl}}(r'')}\frac{\delta \gamma_{\rm{nl}}(r'')}{\delta \rho(r)}\right){\rm{d}}^3{r''}}{\rm{d}}^3{r'}}\\
    =&
    \int{\tau_{\rm{TF}}\frac{\partial F_{\rm{\theta}}^{\rm{NN}}}{\partial \tilde{\xi}_{\rm{nl}}}\int{w(r'-r'')(1-\tilde{\xi}^2(r''))\chi_{\xi}\Big(-\frac{\gamma_{\rm{nl}}(r'')}{\gamma^2(r'')}\frac{1}{3}\frac{\gamma(r'')}{\rho(r'')}\delta(r-r'')}}\\
    &
    + \frac{1}{\gamma(r'')}\frac{1}{3}w(r''-r)\frac{\gamma(r)} {\rho(r)} \Big){\rm{d}}^3{r''}{\rm{d}}^3{r'}\\
    =&
    -\frac{1}{3}\frac{\xi}{\rho}(1-\tilde{\xi}^2)\int{w(r'-r)\tau_{\rm{TF}}(r')\frac{\partial F_{\rm{\theta}}^{\rm{NN}}}{\partial\tilde{\xi}_{\rm{nl}}}{\rm{d}}^3{r'}}\\
    &
    +\frac{1}{3}\frac{1}{\rho^{2/3}}\int{w(r''-r)\frac{1-\tilde{\xi}^2(r'')}{\rho^{1/3}(r'')}\int{w(r'-r'')\tau_{\rm{TF}}(r')\frac{\partial F_{\rm{\theta}}^{\rm{NN}}}{\partial\tilde{\xi}_{\rm{nl}}}{\rm{d}}^3{r'}}{\rm{d}}^3{r''}}.
    \end{aligned}
\end{equation}

Finally, we obtain Eq.~\ref{eq.mpn_p} by combining the five terms derived above.

\section{The FEG limit of Pauli potential}

Given that the kernel function $w(r-r')$ satisfies the condition $\int{w(r-r'){\rm{d}}^3 r'}=0$, in the free electron gas (FEG) limit, all four descriptors become zero. By substituting these equations into Eq.~\ref{eq.mpn_p}, we obtain the following expression:
\begin{equation}
    V_{\theta}^{\rm{MPN}}|_{\rm{FEG}} = \frac{5}{3}\frac{\tau_{\rm{TF}}}{\rho}F_{\rm{\theta}}^{\rm{NN}}|_{\rm{FEG}} = V_{\rm{TF}}F_{\rm{\theta}}^{\rm{NN}}|_{\rm{FEG}},
\end{equation}
where $V_{\rm{TF}}$ represents the potential of TF KEDF.
Therefore, once the FEG limit of Pauli energy $F_{\rm{\theta}}^{\rm{NN}}|_{\rm{FEG}}=1$ is fulfilled, the FEG limit of Pauli potential $V_{\theta}^{\rm{MPN}}|_{\rm{FEG}} = V_{\rm{TF}}$ is automatically satisfied, just as the MPN KEDF does.

\section{The scaling invariance of descriptors}

As explained in the main text, the MPN KEDF subjects to the scaling law 
\begin{equation}
T_{\rm{\theta}}[\rho_{\lambda}] = \lambda^2 T_{\rm{\theta}}[\rho], \rho_{\lambda}=\lambda^3\rho(\lambda r)
\label{eq.scaling}
\end{equation}
thanks to the invariance of the descriptors under the scaling transformation. Specifically, under the scaling transformation $\rho(r) \rightarrow \rho_{\lambda}$, the descriptors $\{\Tilde{p}(r), \Tilde{p}_{\rm{nl}}(r), \Tilde{\xi}(r), \Tilde{\xi}_{\rm{nl}}(r)\}$ transform as $\{\Tilde{p}(\lambda r), \Tilde{p}_{\rm{nl}}(\lambda r), \Tilde{\xi}(\lambda r), \Tilde{\xi}_{\rm{nl}}(\lambda r)\}$.
In the following, we will demonstrate the scaling invariance of these four descriptors.

First, $p(r) = |\nabla \rho(r)|^2 / \Big(2(3\pi^2)^{1/3} \rho^{4/3}(r)\Big)^2$ is invariant under the scaling transformation, because
\begin{equation}
    \begin{aligned}
    p(r)|_{\rho_{\lambda}} 
    &
    = |\nabla \lambda^3\rho(\lambda r)|^2 / \Big(2(3\pi^2)^{1/3} \lambda^4\rho^{4/3}(\lambda r)\Big)^2\\
    &
    = |\nabla_{\lambda r}\rho(\lambda r)|^2 / \Big(2(3\pi^2)^{1/3} \rho^{4/3}(\lambda r)\Big)^2\\
    &
    = p(\lambda r).
    \end{aligned}
\end{equation}
Consequently, $\tilde{p}$ retains this scaling invariance:
\begin{equation}
    \tilde{p}(r)|_{\rho_{\lambda}}
    = \tanh{\Big(\chi_p p(r)|_{\rho_{\lambda}} \Big)}
    = \tanh{\Big(\chi_p p(\lambda r) \Big)}
    = \tilde{p}(\lambda r).
\end{equation}
Hence, we have established the scaling invariance of $\tilde{p}$.

Second, the kernel function $w(r-r')$ will become $\lambda^3 w(\lambda(r-r'))$ under the scaling transformation.
As defined in Eq.~17 of the main text, $w(r-r')$ can be analytical written in reciprocal space as $w(\eta)$, where $\eta = k / 2k_{\tx{F}}, k_{\tx{F}} = (3\pi^2\rho_0)^{1/3}$.
Under the scaling transformation, $k$ becomes $\lambda k$ and $k_{\tx{F}}$ becomes $\lambda k_{\tx{F}}$, so that $\eta$ remains unchanged.
Therefore, the kernel function $w(\eta)$ remains unaffected by the scaling transformation.
However, when we consider $w(r-r') = \int{w(\eta)e^{ik(r-r')}{\rm{d}}^3{k}}$, it undergoes a transformation:
\begin{equation}
    \begin{aligned}
    w(r-r')|_{\rho_\lambda} &= \int{w(\eta)e^{i\lambda k(r-r')}{\rm{d}}^3{(\lambda k)}}\\
    &
    = \lambda^3 \int{w(\eta)e^{i\lambda k(r-r')}{\rm{d}}^3{k}}\\
    &
    = \lambda^3 w(\lambda(r-r')).
    \end{aligned}
\end{equation}
Hence, we observe that $w(r-r')$ transforms as $\lambda^3 w(\lambda(r-r'))$ under the scaling transformation.

Fianlly, $\tilde{p}_{\rm{nl}}, \tilde{\xi}, \tilde{\xi}_{\rm{nl}}$ are invariant under the scaling transformation:
\begin{equation}
    \begin{aligned}
    \tilde{p}_{\rm{nl}}(r)|_{\rho_\lambda} &= \int{\lambda^3 w(\lambda(r-r')) \tilde{p}(\lambda r') {\rm{d}}^3{r'}}\\
    &
    = \int{w(\lambda(r-r')) \tilde{p}(\lambda r') {\rm{d}}^3{(\lambda r')}}\\
    &
    = \tilde{p}_{\rm{nl}}(\lambda r),
    \end{aligned}
\end{equation}
\begin{equation}
    \begin{aligned}
    \tilde{\xi}_{\rm{nl}}(r)|_{\rho_\lambda} &= \tanh{\left(\frac{\int{\lambda^3 w(\lambda(r-r'))\lambda\rho^{1/3}(\lambda r'){\rm{d}}^3{r'}}}{\lambda\rho^{1/3}(\lambda r)}\right)}\\
    &
    = \tanh{\left(\frac{\int{w(\lambda(r-r'))\rho^{1/3}(\lambda r'){\rm{d}}^3{(\lambda r')}}}{\rho^{1/3}(\lambda r)}\right)}\\\\
    &
    = \tilde{\xi}_{\rm{nl}}(\lambda r),
    \end{aligned}
\end{equation}
\begin{equation}
    \begin{aligned}
    \tilde{\xi}_{\rm{nl}}(r)|_{\rho_\lambda} &= \int{\lambda^3 w(\lambda(r-r')) \tilde{\xi}(\lambda r') {\rm{d}}^3{r'}}\\
    &
    = \int{w(\lambda(r-r')) \tilde{\xi}(\lambda r') {\rm{d}}^3{(\lambda r')}}\\
    &
    = \tilde{\xi}_{\rm{nl}}(\lambda r).
    \end{aligned}
\end{equation}
In summary, all of the descriptors $\{\Tilde{p}(r), \Tilde{p}_{\rm{nl}}(r), \Tilde{\xi}(r), \Tilde{\xi}_{\rm{nl}}(r)\}$ are invariant under the scaling transformation.

\section{The details of alloy testing set}

To assess the transferability and stability of the MPN KEDF, we established an alloy testing set comprising 59 alloys sourced from the Materials Project database~\cite{13APL-Jain-MP}.
This set encompasses various combinations, including 20 Li-Mg alloys, 20 Mg-Li alloys, 10 Li-Al alloys, and 9 Li-Mg-Al alloys. 
Comprehensive information regarding these alloys is provided in the following list.

\begin{table*}[htbp]
	\centering
	\caption{Detailed information, including MP IDs, and the components of the alloy testing set, which consists of 20 Li-Mg alloys, 20 Mg-Li alloys, 10 Li-Al alloys, and 9 Li-Mg-Al alloys.}
	\begin{tabularx}{0.99\linewidth}{
			>{\centering\arraybackslash\hsize=0.96\hsize\linewidth=\hsize}X
			>{\centering\arraybackslash\hsize=0.97\hsize\linewidth=\hsize}X
			|>{\centering\arraybackslash\hsize=0.96\hsize\linewidth=\hsize}X
			>{\centering\arraybackslash\hsize=0.97\hsize\linewidth=\hsize}X
			|>{\centering\arraybackslash\hsize=0.96\hsize\linewidth=\hsize}X
			>{\centering\arraybackslash\hsize=0.97\hsize\linewidth=\hsize}X
			|>{\centering\arraybackslash\hsize=0.96\hsize\linewidth=\hsize}X
			>{\centering\arraybackslash\hsize=1.25\hsize\linewidth=\hsize}X}
		\hline\hline
            \multicolumn{2}{c|}{Li-Mg}    &\multicolumn{2}{c|}{Mg-Al}    &\multicolumn{2}{c|}{Li-Al}    &\multicolumn{2}{c}{Li-Mg-Al}\\
		\hline
            MP ID   &Components &MP ID   &Components &MP ID   &Components &MP ID   &Components\\
            \hline
mp-1016222  &1 Li, 3 Mg     &mp-1016233  &3 Mg, 1 Al    &mp-1067     &8 Li, 8 Al    &mp-1015814  &2 Li, 4 Mg, 2 Al\\
mp-1094159  &6 Li, 2 Mg     &mp-1016271  &7 Mg, 1 Al    &mp-1079240  &4 Li, 4 Al    &mp-1023320  &2 Li, 12 Mg, 2 Al\\
mp-1094567  &2 Li, 6 Mg     &mp-1038779  &3 Mg, 3 Al    &mp-1191737  &24 Li, 24 Al  &mp-1023399  &2 Li, 12 Mg, 2 Al\\
mp-1094568  &4 Li, 4 Mg     &mp-1038818  &4 Mg, 4 Al    &mp-1210753  &8 Li, 4 Al    &mp-1028208  &1 Li, 14 Mg, 1 Al\\
mp-1094576  &4 Li, 2 Mg     &mp-1038916  &2 Mg, 6 Al    &mp-1210792  &32 Li, 16 Al  &mp-1028223  &2 Li, 28 Mg, 2 Al\\
mp-1094578  &3 Li, 3 Mg     &mp-1038934  &1 Mg, 1 Al    &mp-1211134  &16 Li, 32 Al  &mp-1185175  &3 Li, 48 Mg, 36 Al\\
mp-1094591  &6 Li, 2 Mg     &mp-1039010  &4 Mg, 4 Al    &mp-16506    &9 Li, 6 Al    &mp-1185468  &2 Li, 34 Mg, 22 Al\\
mp-1094596  &2 Li, 6 Mg     &mp-1039019  &1 Mg, 1 Al    &mp-568404   &9 Li, 4 Al    &mp-973924   &2 Li, 32 Mg, 24 Al\\
mp-1094665  &1 Li, 3 Mg     &mp-1039119  &2 Mg, 6 Al    &mp-975868   &6 Li, 2 Al    &mp-973970   &2 Li, 32 Mg, 24 Al\\
mp-1094670  &1 Li, 3 Mg     &mp-1039141  &1 Mg, 1 Al    &mp-975906   &2 Li, 6 Al    &&\\
mp-1094673  &1 Li, 1 Mg     &mp-1039180  &4 Mg, 12 Al   &&&&\\
mp-1094675  &2 Li, 2 Mg     &mp-1039192  &2 Mg, 4 Al    &&&&\\
mp-1094853  &2 Li, 2 Mg     &mp-1094116  &4 Mg, 8 Al    &&&&\\
mp-1094982  &1 Li, 1 Mg     &mp-1094664  &2 Mg, 2 Al    &&&&\\
mp-1185370  &2 Li, 2 Mg     &mp-1094666  &12 Mg, 4 Al   &&&&\\
mp-1185380  &4 Li, 4 Mg     &mp-1094961  &6 Mg, 2 Al    &&&&\\
mp-1222270  &2 Li, 2 Mg     &mp-1094970  &6 Mg, 2 Al    &&&&\\
mp-865939   &2 Li, 4 Mg     &mp-1094987  &2 Mg, 2 Al    &&&&\\
mp-976139   &3 Li, 1 Mg     &mp-1222013  &1 Mg, 2 Al    &&&&\\
mp-976256   &12 Li, 4 Mg    &mp-2151     &34 Mg, 24 Al  &&&&\\
		\hline\hline
	\end{tabularx}
	\label{tab:s_alloy}
\end{table*}

\section{Random alloys}

\begin{figure}[hbp]
    \centering
    
    \begin{subfigure}{0.48\textwidth}
    \centering
    \includegraphics[width=0.98\linewidth]{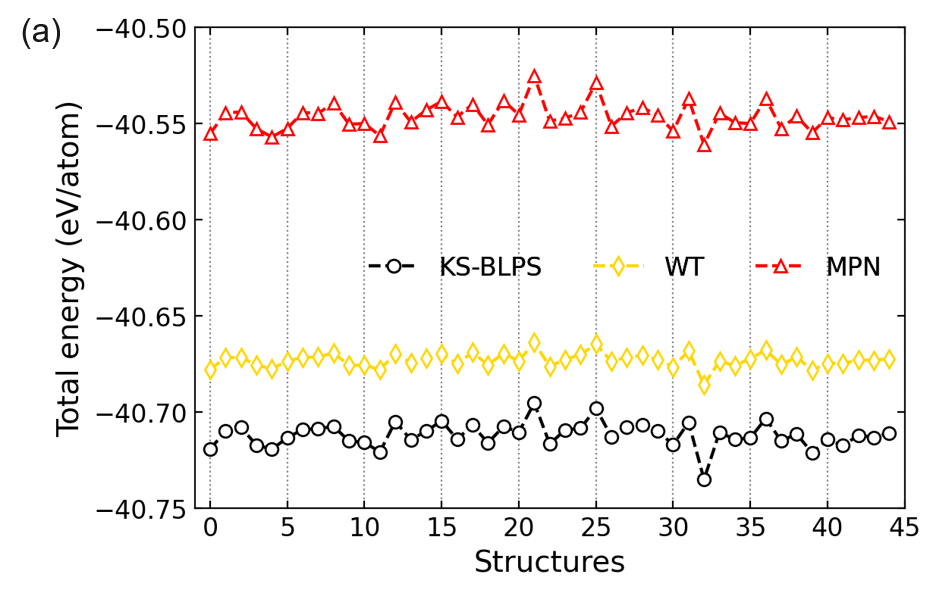}
    \end{subfigure}
    \begin{subfigure}{0.48\textwidth}
    \centering
    \includegraphics[width=0.98\linewidth]{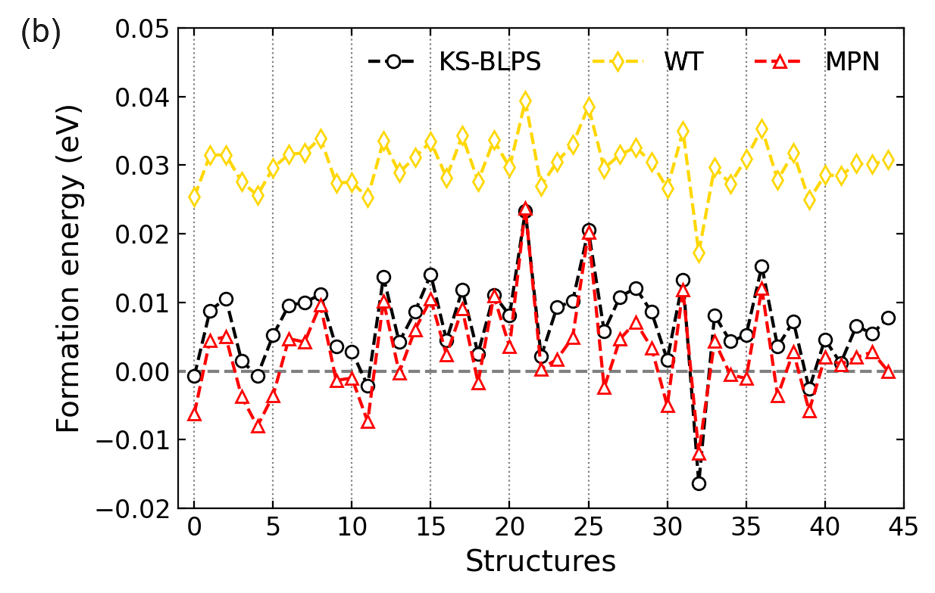}
    \end{subfigure}
    
    \caption{(a) Total energies (in eV/atom) and (b) formation energies (in eV) of 45 Mg-Al alloys, as obtained by WT, MPN KEDFs, and KSDFT with BLPS.}
    \label{fig:SI_alloy}
\end{figure}

\SL{In order to further test the transferability and stability of the MPN KEDF, we have generated 45 hypothetical Mg-Al alloys, and each alloy consists of 16 Al and 16 Mg atoms. In detail, we created a 2×2×2 fcc supercell containing 32 sites and then randomly arranged 16 Al and 16 Mg atoms on these 32 sites. The lattice constant of the supercell is set to 16 Bohr so that the volume per atom is 128 $\rm{Bohr}^3$ per atom.}

\SL{The total energies, as well as formation energies as obtained by KSDFT with BLPS, WT KEDF, and MPN KEDF, are displayed in Fig.~\ref{fig:SI_alloy}.}

\SL{Firstly, in all 45 alloys, MPN KEDF is able to get converged energies, demonstrating its excellent transferability. Second, all of the total energies obtained by WT and MPN KEDF are slightly higher than those obtained by KSDFT, yielding the mean absolute error (MAE) as 0.04 eV/atom and 0.16 eV/atom, respectively. Third, the formation energies obtained by MPN KEDF agree well with those obtained by KSDFT, and the MAE is 0.004 eV, which is much lower than the MAE of WT KEDF, which is 0.023 eV.}

\SL{In conclusion, similar to the phenomenon described in the manuscript, the MPN KEDF gives worse total energies than the WT KEDF, but it yields much more accurate formation energies than WT KEDF, indicating its potential application in the search for the lowest-energy structure of an alloy.}

\bibliography{ML-KEDF}